\documentstyle[emulapj,epsfig]{article}

\font\tenbg=cmmib10 at 10pt

\def \rvecmu{{\hbox{\tenbg\char'026}}}

\def\acc{acc}

\begin{document}


\title{Propagation of Magnetized Neutron Stars
Through the Interstellar Medium.}
\author{O.D. Toropina}
\affil{Space Research Institute, Russian
Academy of Sciences, Moscow, Russia;\\
toropina@mx.iki.rssi.ru}

\author{M.M. Romanova}
\affil{Department of Astronomy, Cornell
University, Ithaca, NY 14853-6801;
romanova@astrosun.tn.cornell.edu}

\author{Yu.M. Toropin}
\affil{Keldysh Institute of Applied
Mathematics,
 Russian Academy of Sciences and
CQG International Ltd.,
\\ 10/5 Sadovaya-Karetnaya,
Build. 1 103006, Moscow, Russia; ytoropin@cqg.com }

\author{R.V.E.~Lovelace}
\affil{Department of Astronomy, Cornell
University, Ithaca, NY 14853-6801;
rvl1@cornell.edu }

\medskip

\begin{abstract}

   This work investigates the propagation of
magnetized, isolated old neutron stars through
the interstellar medium.
     We performed axisymmetric,
non-relativistic magnetohydrodynamic (MHD)
simulations  of the propagation  of
a non-rotating star
with dipole magnetic field aligned with its
velocity through the interstellar medium
(ISM).
Effects of rotation will be discussed in a subsequent work.
     We consider two cases:  (1) where the
accretion radius $R_{acc}$ is comparable  to
the magnetic standoff distance or Alfv\'en radius
$R_A$ and gravitational  focusing is
important;
and (2) where $R_{acc}
<< R_A$  and the magnetized star interacts
with the ISM  as a ``magnetic
plow'',  without significant
gravitational focusing.
     For the first case simulations were
done at a low Mach number ${\cal M}=3$
for a range of values of the magnetic field $B_*$.
     For the second case, simulations
were done for higher Mach numbers, ${\cal
M}=10,$ $30$, and $50$.
   In both cases, the
magnetosphere of the star represents an
obstacle for the flow, and a shock wave
stands in front of the star.
     Magnetic field lines are stretched downwind
from the star and form a
hollow elongated magnetotail.
    Reconnection of the
magnetic field is observed
in the tail which may
lead to acceleration of particles.
   Similar powers are estimated to be
released in the bow shock wave and
in the magnetotail.
The estimated powers are,
however,  below present
detection limits.
Results of our simulations may
be applied to other strongly
magnetized stars, for example,
white dwarfs and magnetic
Ap stars.
   Future more sensitive
observations may reveal long magnetotails
of magnetized stars moving through the
ISM.

\end{abstract}

\keywords{accretion, dipole --- plasmas ---
magnetic fields --- stars: magnetic fields
--- X-rays: stars}

\section{Introduction}

   There are many strongly magnetized stars
moving  through the ISM of our Galaxy.
   One of the most numerous populations
is that of isolated old neutron stars (IONS)
and old magnetars, which are not observed as
radio or X-ray pulsars but which may
still be strongly magnetized.
     There are about $1000$ isolated radio
pulsars observed in the Galaxy.
   The typical age of a radio pulsar is
estimated as $\sim 10^7$ yr
(e.g., Manchester \& Taylor 1977).
    Subsequent to the pulsar stage,
the neutron stars are still
strongly magnetized.
  Pulsar magnetic
fields decay on a longer timescale
than the lifetime of a radio pulsar.
   Thus the number of
magnetized isolated old neutron stars
(MIONS) should be larger than the number of
pulsars.
   Total number of (magnetized and
non-magnetized) isolated old neutron stars
is estimated to be
$10^8 - 10^9$.
   The IONS could be observed
in the solar neighborhood owing to a low-rate
accretion to their surface from the ISM
 (Ostriker, Rees, \& Silk 1970); Schvartsman 1971;
Treves \& Colpi 1991, Blaes \& Madau 1993).
   Many of them may have  strong
magnetic fields, $B\sim 10^9 - 10^{12}~{\rm
G}$ during significant period of their
evolution $\sim 10^8-10^9~ {\rm years}$
(e.g. Livio {\it et al.} 1998, Treves {\it et al.} 2000).

\begin{figure*}[t]
\centering
\epsfig{file=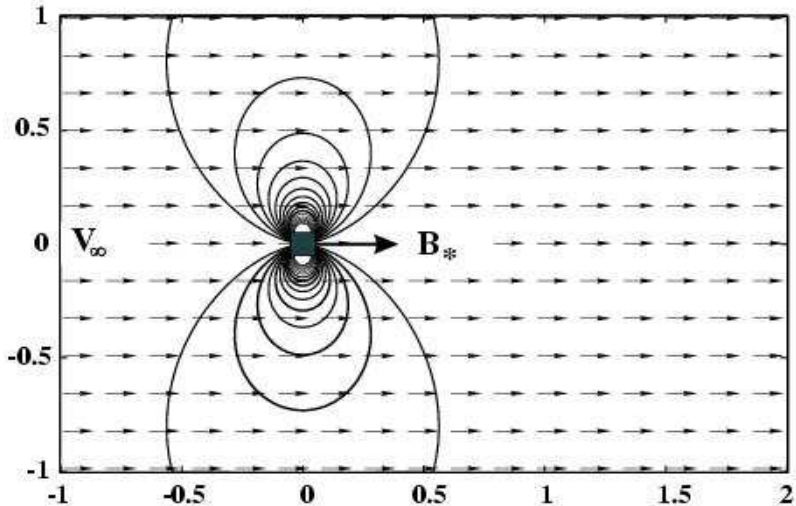,width=4.3in,height=3.in}
\caption{Geometry of the MHD simulation
model.
   The solid
lines are magnetic field lines which are
constant values of the  flux function,
$\Psi(r,z)=$const.
       The $\Psi$
values shown are equally spaced between
$\Psi_{min}=2\times 10^{-5}$ and
$\Psi_{max}=10^{-4}$ in dimensionless
units discussed in \S 3.
 }
\label{Figure 1}
\end{figure*}

    Recently it has been emphasized that some
neutron stars, termed magnetars, may have
anomalously strong magnetic fields at their origin
$B\sim 10^{14}-10^{16}~{\rm G}$ (Duncan \&
Thompson 1992; Thompson \& Duncan 1995
(hereafter TD95); Thompson \& Duncan
1996).
     Magnetars pass through their pulsar stage
much faster than classical pulsars, in
$\sim 10^4$ years (TD95).
    Observations of soft gamma-ray
repeaters (SGRs) and long-period pulsars in
supernova remnants, especially young
supernova remnants (Vasisht \& Gotthelf
1997) support the idea that these objects are
magnetars (Kulkarni \& Frail 1993;
Kouveliotou {\it et al.} 1994).
   The estimated
birthrate of SGRs is $\sim 10\% $ of
ordinary pulsars (Kulkarni \& Frail 1993;
Kouveliotou {\it et al.} 1994, 1999).
   Thus magnetars may constitute a
non-negligible percentage of IONS.

  MIONS and magnetars typically
move supersonically
through the ISM and have
extended magnetospheres.
   Two main regimes are
possible:
    In the first, the Alfv\'en radius
$R_A$ is much smaller than gravitational
accretion radius $R_{acc}$, so that matter
is gravitationally attracted by the star and
direct accretion to a star is possible
(e.g. Hoyle and Lyttleton 1939; Bondi 1952;
Lamb, Pethick, \& Pines 1973; 
Bisnovatyi-Kogan \& Pogorelov 1997).
   In the second regime, the
magnetic standoff distance or
Alfv\'en radius $R_A$ is larger
than accretion radius and the magnetosphere
interacts with the ISM  without
gravitational focusing.
   This case we term the
``magnetic plow'' regime (this corresponds 
to ``georotator" regime
in Lipunov, 1992).
     This is the regime for fast moving
MIONS and  magnetars.
   Neither of these regimes was investigated
numerically in application to magnetized
star propagation through the ISM.
    Most of simulations of this type
were done to model the interaction
of the Earth's
magnetosphere with the Solar wind
(e.g., Nishida, Baker \& Cowley 1998),
 where parameters of the problem were fixed
by the Solar wind and Earth's magnetic
field.

    In this paper we
investigate the supersonic motion
of magnetized stars through the ISM
where a wide range of
physical parameters is possible.
     We investigate the physical process of
interaction of magnetospheres with the ISM
and estimate the possible observational
consequences of such interaction.
     In \S 2 we
estimate the important physical parameters, and
in \S 3  we describe the numerical model.
    In \S 4 we summarize results of simulations for
$R_A \lesssim  R_{acc}$ and for a small Mach number,
${\cal M}=3$.
    In \S 5 we discuss results of simulations
in the ``magnetic plow'' regime.
     In \S 6 we
discuss possible observational consequences
of our results.
    In \S 7 we give a brief summary.

\begin{figure*}[b]
\centering
\epsfig{file=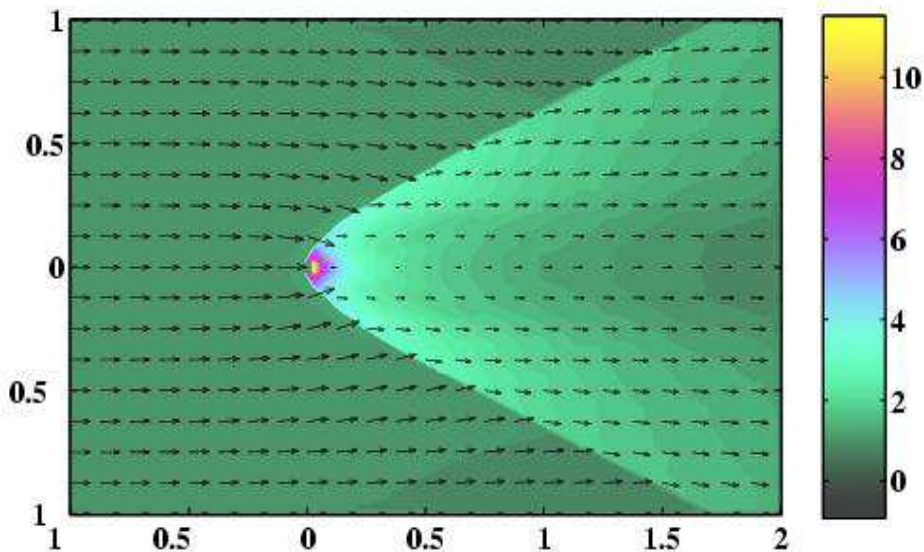,width=4.9in,height=3.in}
\caption{ Results of hydro simulations
 of accretion to a non-magnetized star at
Mach number
${\cal M}=3$. The background and contours
represent density. Arrows represent
velocity vectors.}
\label{Figure 2}
\end{figure*}

\section{Physical Model}

  After the radio pulsar stage,
neutron stars are
still strongly magnetized
and rotating objects.
    This work treats the
motion of a non-rotating
magnetized star through the
interstellar medium.
    Treatment of the motion
of a rotating star through the
ISM is discussed by
Romanova {\it {\it et al.}} (2001).

   A non-magnetized star moving
through the ISM  captures
matter gravitationally from the accretion
radius (e.g. Shapiro \& Teukolsky 1983),
$$
 R_{acc}= {2GM \over c_s^2+v^2}\approx
9.4\times 10^{11}{ M_{1.4}  \over v_{200}^{2}}~{\rm
cm}~,
\eqno(1)
$$
where $v_{200}\equiv v/(200~{\rm km/s})$
is the normalized velocity of the star,
$c_s$ the sound speed of the undisturbed
ISM, and
$M_{1.4} \equiv M/(1.4 M_\odot)$ is the normalized mass
of the star.
   The mass accretion rate
at high Mach numbers ${\cal
M}\equiv v/c_s>>1$ was derived by Hoyle and
Lyttleton (1939),
$$
\dot M_{HL}=4\pi (GM)^2 {\rho \over v^{3}}\approx
9.3\times 10^7 {n \over v_{200}^{3}} M_{1.4}^2~{\rm
\frac{g}{s}}~,
\eqno(2)
$$
where $\rho$ is the mass-density of the ISM and
$n=n/{1~{\rm cm}^{-3}}$ is the normalized number
density.
    For arbitrary ${\cal M}$ a general
formulae was proposed by Bondi (1952),
$$
\dot M_{BHL} = \pi R_{acc}^2 \rho v =4\pi
\alpha (GM)^2 {\rho \over (v^2+c_s^2)^{3/2}}~,
\eqno(3)
$$
where  the coefficient $\alpha$ is of
order unity (e.g., Bondi proposed
$\alpha=1/2$, see also Ruffert 1994a,b;
Pogorelov, Ohsugi, \& Matsuda 2000).

   For the case of a moving magnetized
star,  the standoff distance where the
inflowing ISM is stopped by the star's
$B-$field is referred to as the Alfv\'en
radius $R_A$.
   For a relatively weak
stellar magnetic field
$R_A \ll R_{acc}$, and in this limit
of ``gravitational accretion''
denote the Alfv\'en radius as $R_{Ag}$.
   The accretion flow  becomes
spherically symmetric inside $R_{acc}$,
and one finds
$$
  R_{Ag}= \bigg(\frac{B_*^2 R_*^6}
{\sqrt{2GM} \dot M}\bigg)^{2/7}~{\rm cm}~,
\eqno(4)
$$
(e.g., Lamb, et al. 1973, Lipunov 1992),
where $B_*$ is the magnetic field at the
surface of the star of radius $R_*$
and $\dot{M}$ is the accretion rate.
  If a magnetized star accretes matter with the
same rate as a non-magnetized star, $\dot M
= \dot M_{BHL}$, then the Alfv\'en radius
is
$$
    R_{Ag}\approx 1.2\times 10^{11}\frac{B_{12}^{4/7}
R_6^{12/7} v_{200}^{6/7}} {M_{1.4}^{5/7}
n^{2/7}}~ {\rm cm}~,
\eqno(5)
$$
which is  about $R_{acc}/8$ for the
adopted reference parameters.  Here,
$B_{12}\equiv B_*/10^{12}$G and
$R_6 \equiv R_*/10^6$cm.

   However, there is reason
to believe that a {\it magnetized} star
accretes matter at a {\it lower} rate than
a non-magnetized star for the same $v, ~c_s,$
and $M$.
    Our study
of spherical Bondi accretion
has shown that the
magnetized star accretes at a lower rate
than the same
non-magnetized star (Toropin {\it et al.}
1999, hereafter T99).
   The magnetosphere
acts as an  obstacle for the flow,
thus decreasing the rate of spherical
accretion compared to the
Bondi rate
$\dot{M}_B=
4\pi\alpha(GM)^2\rho/c_s^3$.
     Equations (28) and (32) of T99
correspond to the approximate dependence
$$
\frac{\dot M}{\dot M_{B}}\approx
\bigg(\frac{R_*}{R_{Ag}^{th}}\bigg)^{7/4}~,
\eqno(6)
$$
for $R_{Ag}^{th}/R_*$ in the range $1-10$,
where $R_{Ag}^{th}$ is
given by equation (4) with $\dot{M}=\dot{M}_B$.
    Thus, for a larger  $R_{Ag}^{th}$, $\dot{M}$
is smaller and the actual Alfv\'en radius given
by equation (4) is larger.

   Equation (6)
was deduced from simulations
at small values of $R_{Ag}/R_*$ and
therefore it cannot be reliably
extrapolated to very
large values of this ratio.
    Instead we can
write in general $\dot{M}={\cal K}\dot{M}_B$
where ${\cal K} \leq 1.$
   Then we find that the actual Alfv\'en
radius is
$ \tilde R_{Ag} = R_{Ag}^{th} {\cal K}^{-2/7}$.
    The two
 radii, $R_{acc}$ and $\tilde R_{Ag}$ are
equal at $ {\cal K}\approx 10^{-3}$ for our
reference parameters.
     It is not known
whether  accretion can be so strongly
inhibited at such  small values of ${\cal K}$.

\begin{figure*}[t]
\centering
\epsfig{file=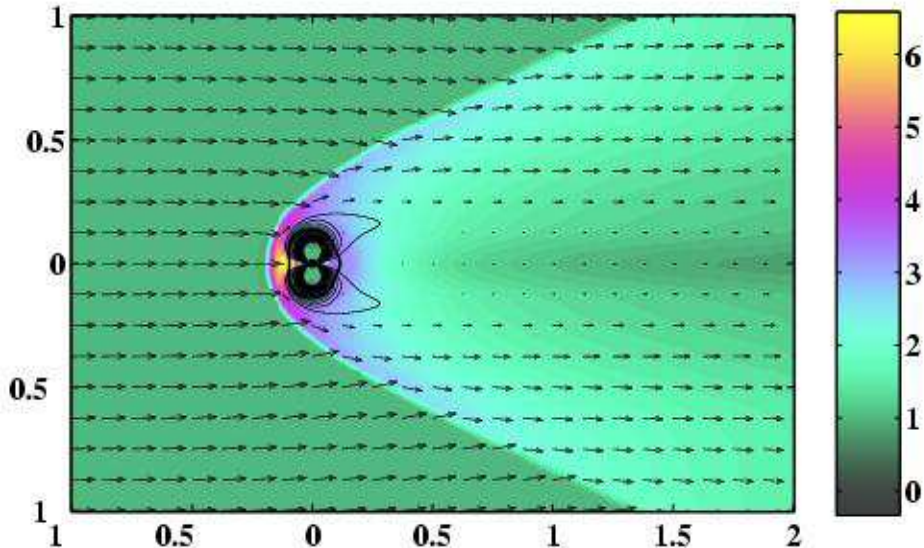,width=4.9in,height=3.in}
\caption{ Results of simulations of
accretion to a magnetized star with magnetic
field $B_*=3.5$ at Mach number
${\cal M}=3$. Poloidal magnetic field lines
and velocity vectors $\bf v$ are shown. The
background represents density.
  The thick  line represents the Alfv\'en
surface.}
\label{Figure 3}
\end{figure*}

\begin{figure*}[b]
\centering
\epsfig{file=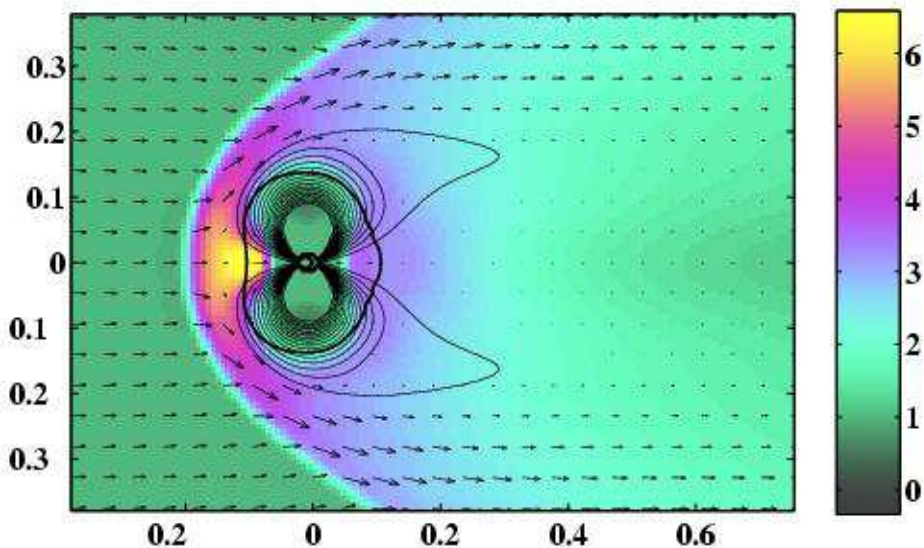,width=4.9in,height=3.in}
\caption{ Same case as Figure 3, but the inner
region is shown at higher resolution. Arrows
show matter flux vectors
$\rho {\bf v}$.
  The thick  line represents the Alfv\'en
surface.}
\label{Figure 4}
\end{figure*}

  Magnetars have significantly
stronger  magnetic fields than
typical radio pulsars, and consequently
most of them are in the ``magnetic
plow'' regime.
  Comparison of equations (1) and (5)
shows that  $R_{acc} \leq \tilde{R}_A$ if
$$
B_*\ge 3.7\times 10^{13}\frac{{\cal K}^{1/2} M_{1.4}^3
n^{1/2}}{R_6^3 v_{200}^5}~{\rm G}~.
\eqno(7)
$$
Thus, even for ${\cal K}=1$ and
$v\approx 200~{\rm km/s}$, magnetars are in
the magnetic plow regime.

   In the ``magnetic plow'' regime,
the Alfv\'en radius $R_{Ap}$ follows from
the balance of the
 magnetic pressure of
the star $B^2/4\pi = B_*^2(R_*/R)^6$
against the ram pressure of the ISM
which is $\rho v^2$ for Mach numbers
${\cal M}\gg1$.
Thus
$$
R_{Ap}= R_* \bigg(\frac{B_*^2}{4\pi\rho
v^2}\bigg)^{1/ 6}\approx 2.2\times 10^{11} R_6
\bigg(\frac{B_{12}^2}{4\pi n
v_{200}^2}\bigg)^{1 / 6} ~{\rm cm}~.
\eqno(8)
$$
The magnetic field strength at this
distance from the star is
$$
B_A=(4\pi \rho)^{1/2} v \approx 9.2\times
10^{-5} n^{1/2} v_{200}~{\rm G}~.
\eqno(9)
$$
   At the boundary between the
gravitational and magnetic plow regimes,
equations (1), (4), and (8) coincide.
   We mention here  the important
influence of rotation
of the star.
  Due to the rotation the magnetic
field decreases with distance $\propto 1/r$
at large distances
rather than $\propto 1/r^3$ so
that the Alfv\'en radius is
much larger than one described by equation (8) (Romanova
{\it et al.} 2001).

\begin{figure*}[t]
\centering
\epsfig{file=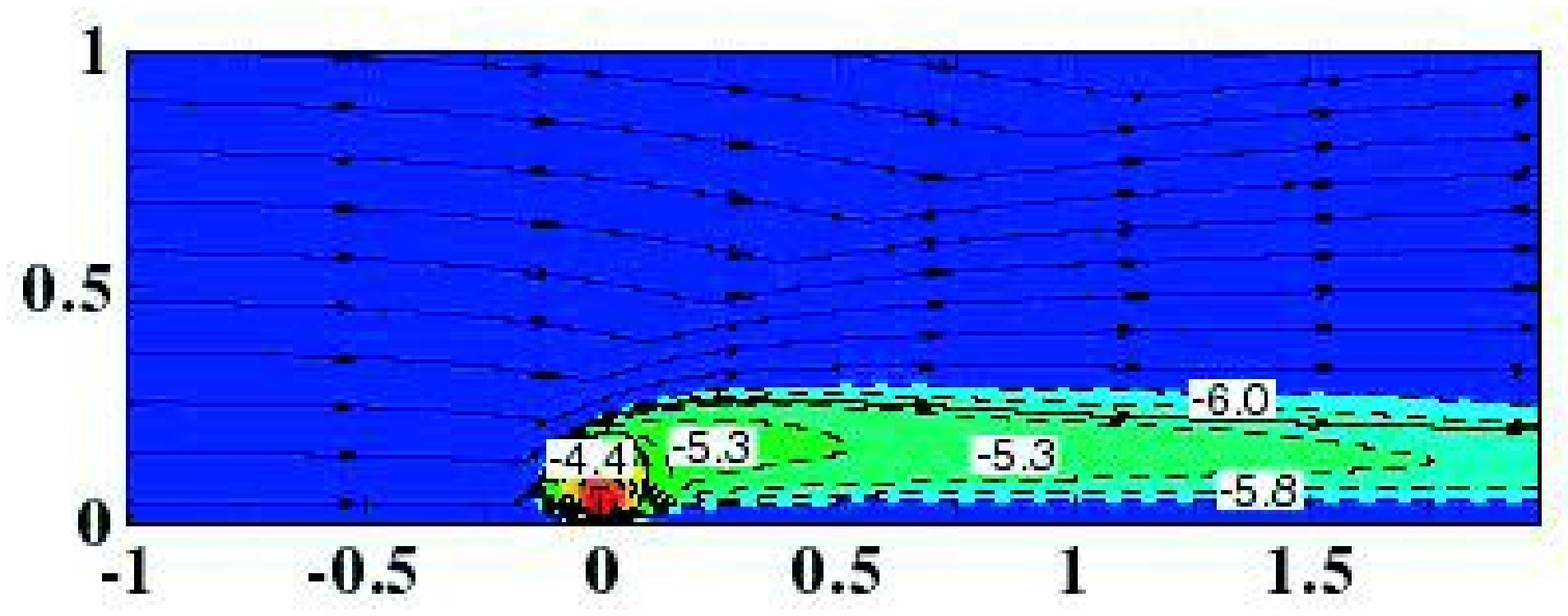,width=4.5in,height=1.7in}
\caption{
 Same case as Figure 3, but the
streamlines  of matter flux $\rho {\bf v}$
are shown.
    Background and dashed lines
represent logarithm of magnetic flux which
is equally spaced between
$\log_{10}\Psi =-6$ and $\log_{10} \Psi = -4$.
  The minimum
value of
$\Psi$ is smaller than that in Figure 3.
Numbers show the value of logarithm of $\Psi$.
}
\label{Figure 5}
\end{figure*}

\begin{figure*}[b]
\centering
\epsfig{file=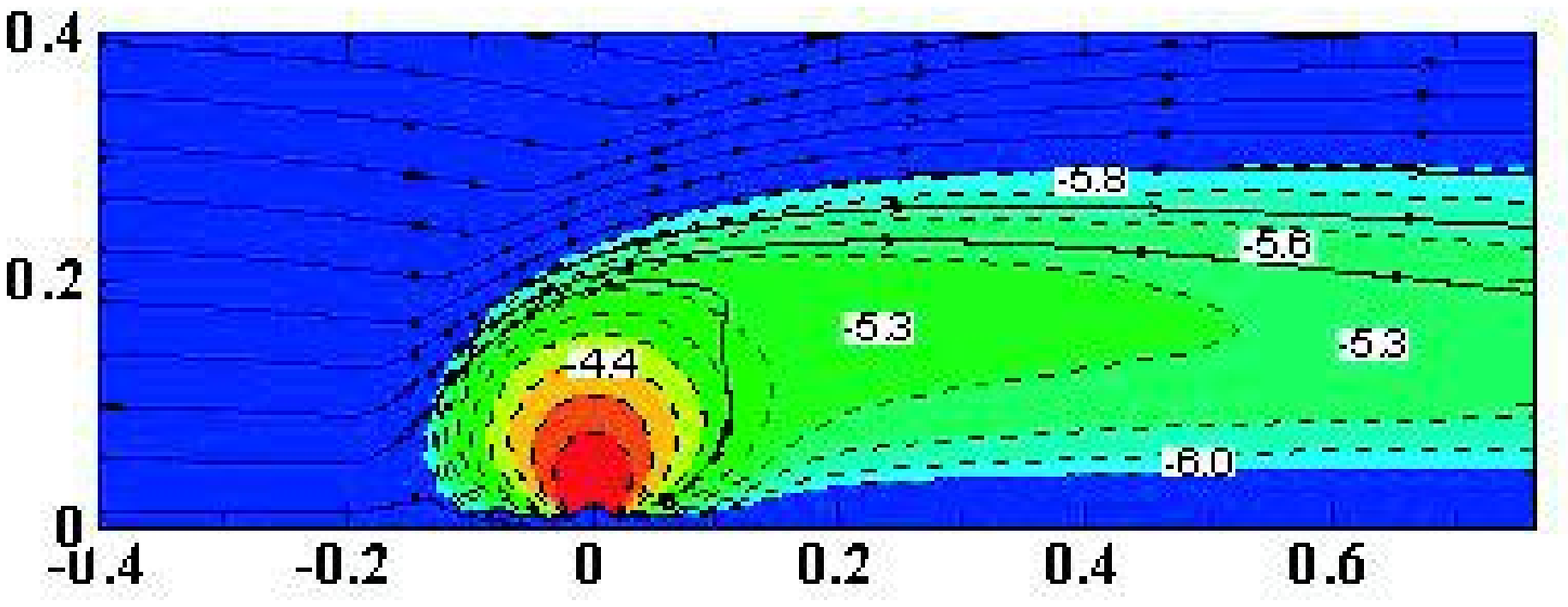,width=4.5in,height=1.7in}
\caption{
 Same case as Figure 5, but the inner region is
shown at higher resolution.}
\label{Figure 6}
\end{figure*}

  The velocity distribution
of MIONS and magnetars
is unknown, but it is expected
to be similar to that of radio
pulsars.
   Pulsars have a wide range of velocities,
$10~{\rm km/s} \le {\bar v} \le 1500~{\rm km/s}$,
with the peak of the distribution at $\bar
v \approx  175~{\rm km/s}$ (Cordes \&
Chernoff 1998).
   Some authors give a smaller value,
${\bar v}\approx 100~{\rm km/s}$ (Narayan \&
Ostriker 1990) while others give
larger values,
 ${\bar v}\approx 250 - 300 ~{\rm km/s}$
 (Hansen \& Phinney 1997),
 ${\bar v}\approx 500~ {\rm km/s}$
(Popov {\it et al.} 2000).
 For temperatures of the ISM, $T\approx
10^4~{\rm K}$, the sound speed of gas is
$c_s\approx \sqrt{\gamma k_B T/\bar{m}}\approx
11.7~ {\rm km/{s}}$, where $\bar{m}\approx m_p$
is the mean particle mass, $k_B$ is
Boltzmann's constant, and $\gamma$ is the
usual specific heat ratio.
  Thus, the Mach number
of radio pulsars is in the range
${\cal M}=v/c_s \sim 1-150$ with most
pulsars having
${\cal M} \sim 10-50$.
  The accretion
radius $R_{acc}$  depends strongly on the
velocity of the star $v$, which  may change
the ratio between $R_{acc}$ and $R_A$ and
correspondingly the regime of accretion.
  For
example, very fast MIONS with $v\sim 1000~
{\rm km/s}$ have much smaller accretion
radii than slow ones, and may have $R_A >>
R_{acc}$ for wide range of magnetic fields.

    It is clear from the range of  surface magnetic
fields of MIONS and magnetars and the range
of their velocities  that different regimes are
 possible:   (1) The regime of
gravitational accretion, $R_{acc}>>R_A$;
(2) The intermediate regime,
 $R_{acc}\sim R_A$;
 and (3) The ``magnetic plow'' regime,
$R_{acc}<<R_A$.
   In this paper, we present
results for regimes (2) and (3), which are
characterized by formation of extended
magnetotails.
  Regime (1) will be investigated
in a future work.
   Below, we present numerical
model and results of simulations, and
return to discuss  physical model further
 in \S 6,
where the possible observational
consequences are  considered.

\section{Numerical Model}

To investigate the interaction
of a magnetized star with
the ISM we use an
axisymmetric resistive MHD code and arrange
dipole so that its axis is aligned with the
matter flow (see Figure 1).
    The code uses
a total variation diminishing (TVD) method
(Savelyev {\it et al.} 1996; Zhukov {\it et al.}
1993).
The code was used earlier for
a study of spherical Bondi accretion
to a star with dipole magnetic field (T99).

   We used a cylindrical coordinate system
$\left(r,\phi,z\right)$ with its origin at
the star's center.
   The $z$-axis is parallel to the velocity
of the ISM at large distances ${\bf v}_\infty$.
  The dipole magnetic
moment of the star ${\rvecmu}$ is parallel
or antiparallel to the $z-$axis.
  Axisymmetry is
assumed so that ${\partial}/{\partial \phi}=0$
for all scalar variables.
   We solve for the vector potential ${\bf A}$
so that the magnetic field
${\bf B}={\bf \nabla}\times {\bf A}$
automatically satisfies
${\bf \nabla}\cdot{\bf B}=0$.

  The flow is described by the
resistive MHD equations,
$$
  {\partial \rho \over
  \partial t}+
  {\bf \nabla}{\bf \cdot}
\left(\rho~{\bf v}\right)=0~,
\eqno(10)
$$
$$
\rho\left(\frac{\partial }{\partial t}+
\left({\bf v}{\bf \cdot}{\bf \nabla}\right)
\right){\bf v} =
 -{\bf \nabla}p+
{1\over c}{\bf J\times B} +{\bf
F}^g~,
\eqno(11)
$$
$$
  \frac{\partial {\bf B}} {\partial t}=
  {\bf \nabla}{\bf \times}
\left({\bf v}{\bf \times} {\bf B}\right)
  +
  \frac{c^2}{4\pi\sigma}
\nabla^2{\bf B} ~,
\eqno(12)
$$
$$
  \frac{\partial (\rho\varepsilon)}
{\partial t}+
  {\bf \nabla}\cdot \left(\rho
\varepsilon{\bf v}\right)=
  -p\left({\bf \nabla}{\bf \cdot} {\bf v}\right)
+{1 \over \sigma}{\bf J}^2 ~.
\eqno(13)
$$
  The variables have their usual
meanings.
   The equation of state is
$p=\left(\gamma-1\right)\rho
\varepsilon$, with
$\gamma$ the specific heat ratio.
   In the simulations presented
here $\gamma=5/3$.
   The equations incorporate Ohm's law
${\bf J}=\sigma({\bf E}+{\bf v}
 \times {\bf B}/c)$, where $\sigma$ is the
electrical conductivity.
   The corresponding
magnetic diffusivity
$\eta_m \equiv c^2\!/(4\pi\sigma)$
is taken to be a constant.

\begin{figure*}[t]
\centering
\epsfig{file=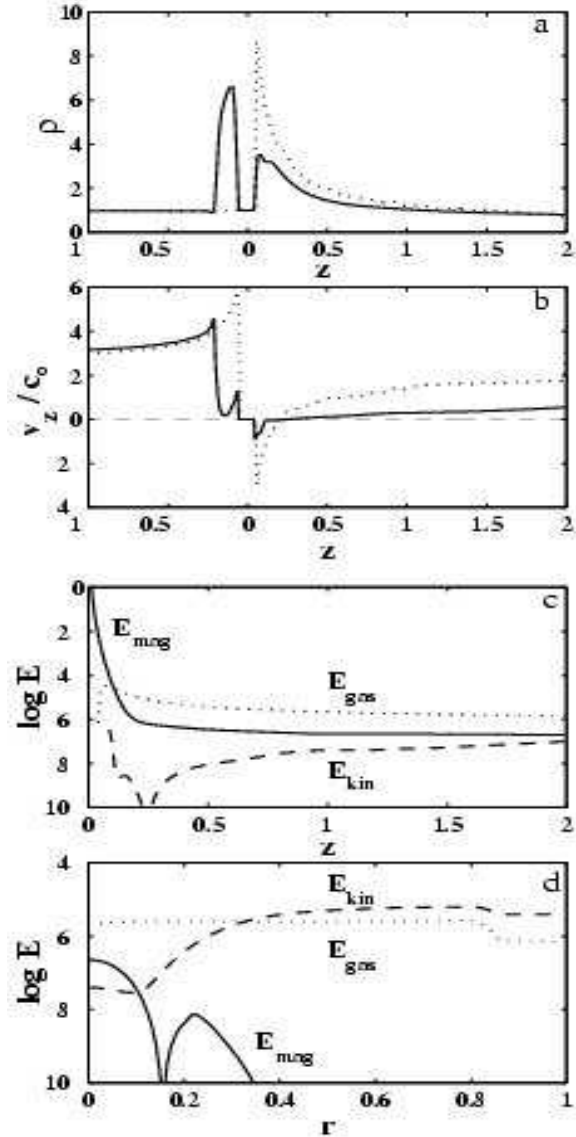,width=3in,height=6in}
\caption{ Panel (a) shows the density
and panel (b) the velocity variation
along the
$z-$ axis.
    Panel (c) shows the
energy-density variation along the
$z-$ axis, and panel (d)
shows its variation with $r$  at
$z=1$.
Here,
$E_{mag}-$is the magnetic energy-density,
$E_{kin}-$ is the kinetic
energy-density, and $E_{gas}$ is the
thermal energy-density.
   For the case shown, $B_*=3.5$
and ${\cal M}=3$. Dotted lines on panels (a,b)
correspond to hydrodynamic simulations.
}
\label{Figure 7}
\end{figure*}

  The simulations were done inside a
``cylindrical box,''
$(Z_{min}\le z\le Z_{max}, 0\le r\le
R_{max})$.
   A uniform $(r,z)$ mesh was used
with size $N_R\times N_Z.$
    The magnetized
star   was
represented  by a small
cylindrical box  with dimensions
$R_*<<R_{max}$ and $|Z_*|<<Z_{max}$,
which constitutes the ``numerical star.''
   In equation (11)
the gravitational force is due
to the star,
${\bf F}^{g} = -GM\rho{\bf
R}/\!R^3$.
    The gravitational force was
smoothed for distances inside the region
 $r = |z| = 0.25 R_*$ which does
not influence the computational results
outside of the numerical star.

A point dipole magnetic field
${\bf B} =[3{\bf R}\left({\rvecmu}\cdot {\bf R}
\right)-R^2 \rvecmu ]/{R^5}$ with
vector-potential
${\bf A}={{\rvecmu}\times{\bf R}}/{R^3}$
 was arranged inside numerical star
at the radii: $r > 0.25 R_*$.
    This
dipole field differs from that used in T99
where a small but
finite size ``current'' disk was used
to produce the dipole field.
   A similar model of the field was used
by Hayashi {\it et al.} (1996),
 Miller \& Stone (1997),
 Goodson {\it et al.} (1997).

  The vector potential was fixed inside the
numerical star and at its surface
during the simulations.
   These conditions follow from the
${\bf E}$ and ${\bf B}$
boundary conditions on the surface of
the perfectly conducting star and protect
the  magnetic field against numerical decay
(T99).
  The hydrodynamic variables $\rho,$ $v_r,$
$v_z,$ and
$\varepsilon$ were fixed at the surface of
the numerical star.
   These conditions are similar to the standard
``vacuum'' conditions adopted in hydrodynamic
simulations (e.g., Ruffert 1994a,b).
   However, the vacuum is not made
too strong
because of the difficulty of handling low
densities in MHD simulations.
   We discuss the boundary conditions
on the numerical star further
in \S 4.1.
   We tested
the influence of the numerical star shape on
our simulation results.
   Namely, we created an
approximation of a sphere on
rectangular grid and compared it with the
cylindrical star and observed that the
difference in the shapes has an insignificant
influence on our results.

\begin{figure*}[t]
\centering
\epsfig{file=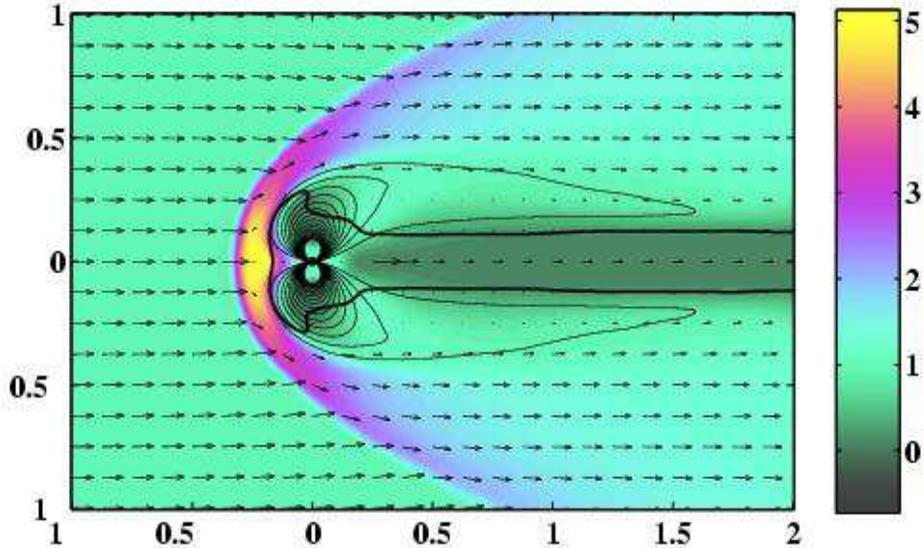,width=4.9in,height=3.in}
\caption{Results of simulations of motion
of a magnetized star with
magnetic field $B_*=14$ through the ISM with
 Mach number
${\cal M}=3$.
   Magnetic field lines and
velocity vectors are shown.
   The background
represents the density.
    The thick  line indicates the Alfv\'en
surface.}
\label{Figure 8}
\end{figure*}

\begin{figure*}[b]
\centering
\epsfig{file=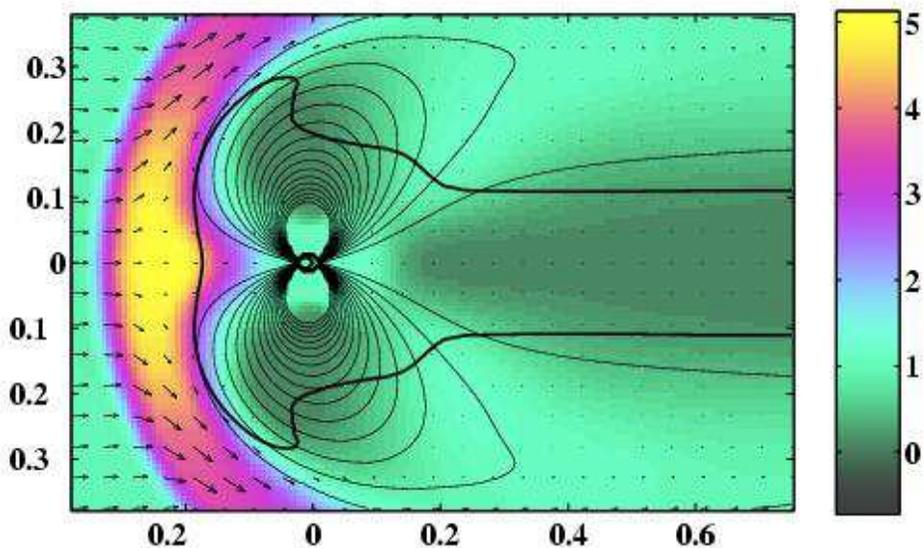,width=4.9in,height=3.in}
\caption{ Same run as  Figure 8,
but the inner
region is shown at higher resolution.
Arrows show matter flux vectors $\rho \bf v$.
  The thick line
represents the Alfv\'en surface.}
\label{Figure 9}
\end{figure*}

    We put the MHD equations in
dimensionless form using the following
scalings:
   The characteristic length is taken
to be the Bondi radius,
$R_B={GM}/c_{s\infty}^2$,
where $c_{s\infty}$ is the sound
speed in the undisturbed ISM.
   Temperature is measured in units
of $T_\infty$, and density
in units of $\rho_\infty$.
    The magnetic field is measured
in units of the reference magnetic field $B_0$.
   A reference speed
is the Alfv\'en velocity
corresponding to a reference magnetic field
$B_0$ and density $\rho_\infty$,
$v_0 \equiv {B_0}/\sqrt{4 \pi
\rho_\infty}$.
    Time is measured in units of
$t_0=(Z_{max} - Z_{min})/v_\infty$, which is the
crossing time of the computational region
in the absence of a star.
    After reduction to  dimensionless form,
the MHD equations (10) - (13) involve
three dimensionless parameters,
$$
\beta \equiv \frac{8\pi p_\infty}{B_0^2}~, \quad\quad
g \equiv \frac{G M}{R_B v_0^2} =
\frac{1}{2} \gamma \beta~,
\eqno(14)
$$

$$
\tilde{\eta}_m \equiv {\eta_m \over R_B
v_0} = {1\over Re_m}{~,}
\eqno(15)
$$
where $\tilde{\eta}_m$ is the dimensionless
magnetic diffusivity, and
$Re_m$ is the
magnetic Reynolds number.
    Note that the
first two parameters are dependent because
of our choice of the length scale $R_B$.

\begin{figure*}[t]
\centering
\epsfig{file=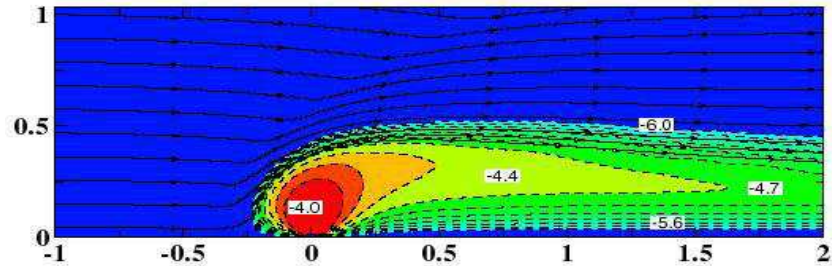,width=4.5in,height=1.7in}
\caption{ Same case as Figure 8, but the
streamlines (solid lines) of
matter flux $\rho {\bf v}$ are
shown.
  The background represent
logarithm of magnetic flux which is equally
spaced between
$\log_{10} \Psi =-6$ and $\log_{10} \Psi = -4$.
The numbers indicate
the logarithm of $\Psi$.}
\label{Figure 10}
\end{figure*}

    The external boundaries of the
computational region were treated
as follows.
   Supersonic inflow with Mach
number ${\cal M}$ was specified
at the upstream boundary
$(z=Z_{min}, 0\le r \le R_{max})$.
   At the downstream boundary
$(z=Z_{max}, 0\le r \le R_{max})$, a ``free
boundary'' condition was applied,
$\partial/\partial{\bf n}=0$.
   Inflow of matter from  this boundary
into the computational region was
forbidden.
   At the
cylindrical boundary
$(Z_{min}\le z\le Z_{max}, r=R_{max}$, we
used the free boundary conditions and in some cases
forbid inflow to the computational region.
We observed that the result is very
similar in both cases.
    We checked the
influence of external boundary conditions by
performing test simulations at different sizes
of the computational region.

    The size of the computational region for
most of the simulations
was  $R_{max}=2 R_{B}=2$,
$Z_{min}=-R_{max}=-2$, and $Z_{max}=2R_{max}=4$.
   The grid $N_R\times N_Z$ was
$257\times 769$ in most of cases.
    The radius of the numerical star
was $R_*=0.05{}R_{B}=0.05$ in most cases,
but test runs were also done for
$R_*=0.02$.
   A number of different
values of $R_*$
were investigated
in the purely  hydrodynamic
simulations (see \S 4.1).

   For most of our simulation runs
$\beta=10^{-6}$.  Therefore,
our reference magnetic field
$B_0 =\sqrt{8\pi p_\infty/\beta}$
is also fixed since $p_\infty$ is fixed.
   Thus a useful measure of the
strength of the magnetic field is the
ratio of the maximum  value of the $z-$component
of the field at the point  $r=0.25 R_*$ and $z=0$
to  $B_0$.
    We denote this dimensionless field as $B_*$.
      We performed simulations for a range
of values of $B_*$.
      The magnetic diffusivity was taken to be
$\tilde{\eta}_m=10^{-6} $ in most of runs, but
the dependence of our solutions  on $\tilde{\eta}_m$
is discussed in \S 5.2.

\begin{figure*}[b]
\centering
\epsfig{file=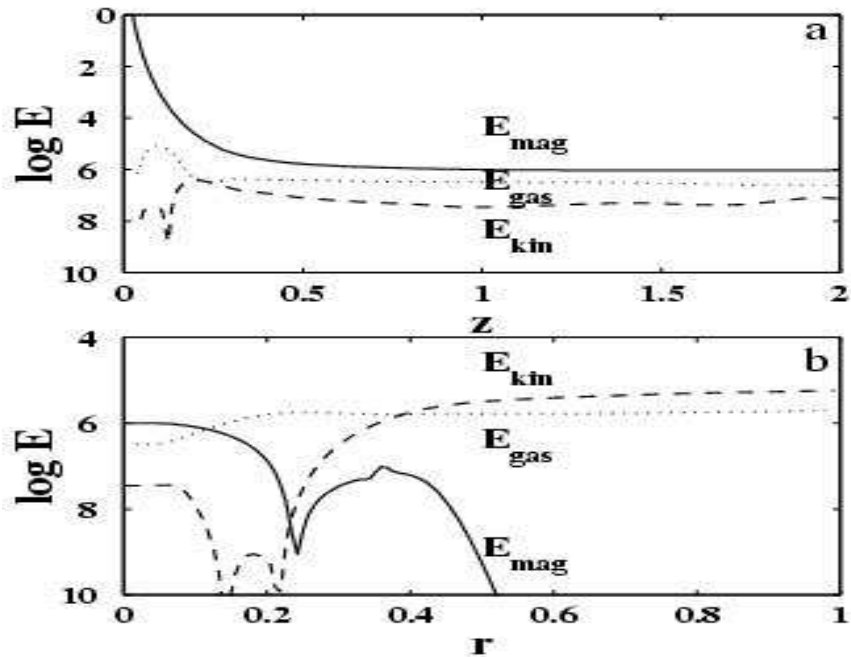,width=4.5in,height=3.5in}
\caption{Panel (a) shows
the energy-density variation
along the tail.
    Panel (b) shows the variation
across the tail at $z=1$.
   Both panels are for $B_*=14$, ${\cal M}=3$.
Here, $E_{mag}$ is the magnetic energy-density,
$E_{kin}-$ the kinetic
energy-density, and $E_{gas}$ the thermal
energy-density. }
\label{Figure 11}
\end{figure*}

    Initially, at  $t=0$ the
magnetic field of the star
 is a dipole field.
   The density and flow velocity are
homogeneous in the  simulation region:
$\rho=\rho_\infty$ and $v=v_\infty$ (see Figure 1).
   We investigate subsequent
evolution and follow the evolution
as long as it is needed to reach
stationarity or quasi-stationarity.
   This is typically several dynamical time-scales.

\begin{figure*}[t]
\centering
\epsfig{file=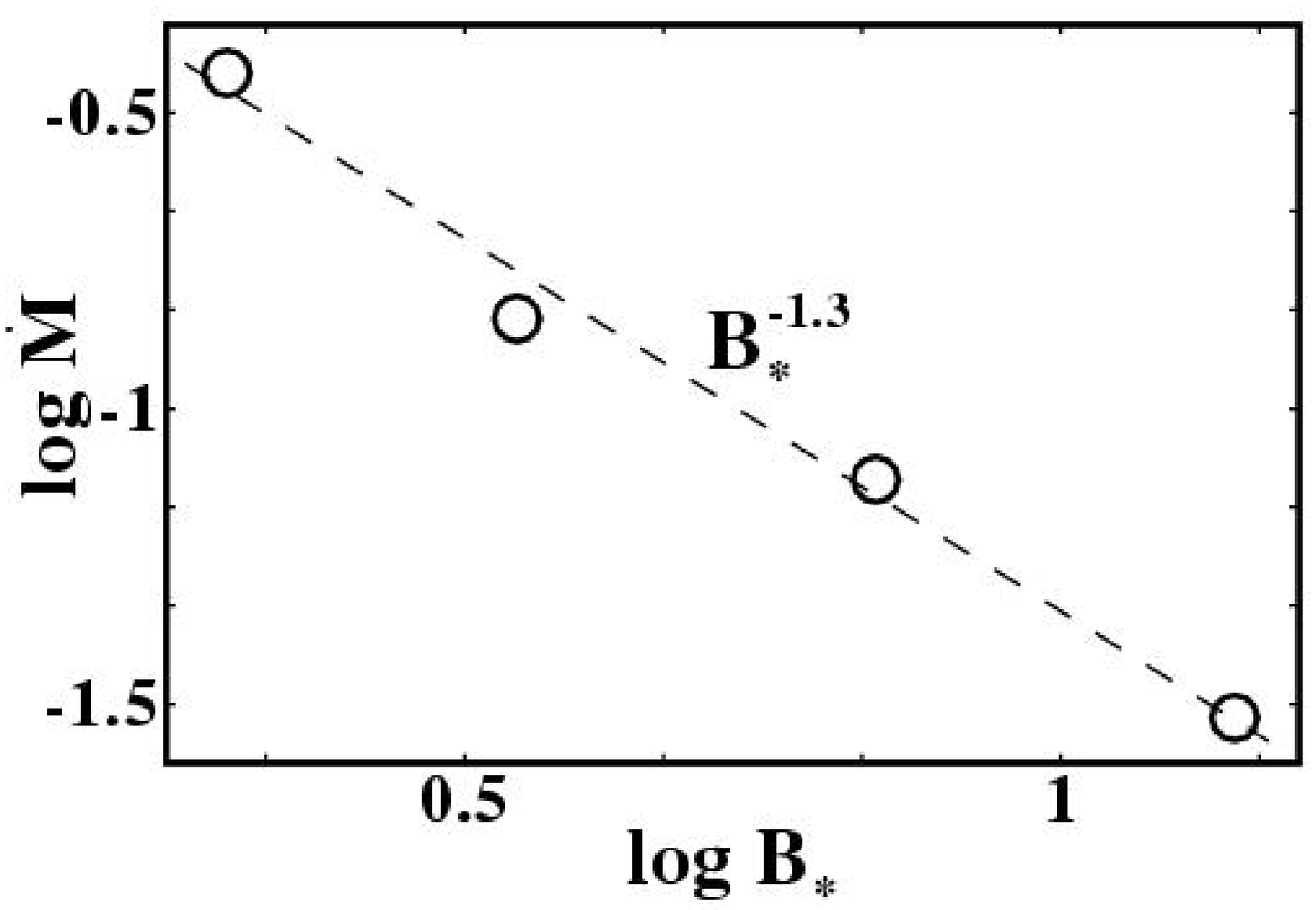,width=4.5in,height=3.5in}
\caption{ Dependence of the accretion rate to a
star on the surface
 magnetic field $B_*$ for
a star moving at Mach number ${\cal M}=3$.}
\label{Figure 12}
\end{figure*}

\begin{figure*}[b]
\centering
\epsfig{file=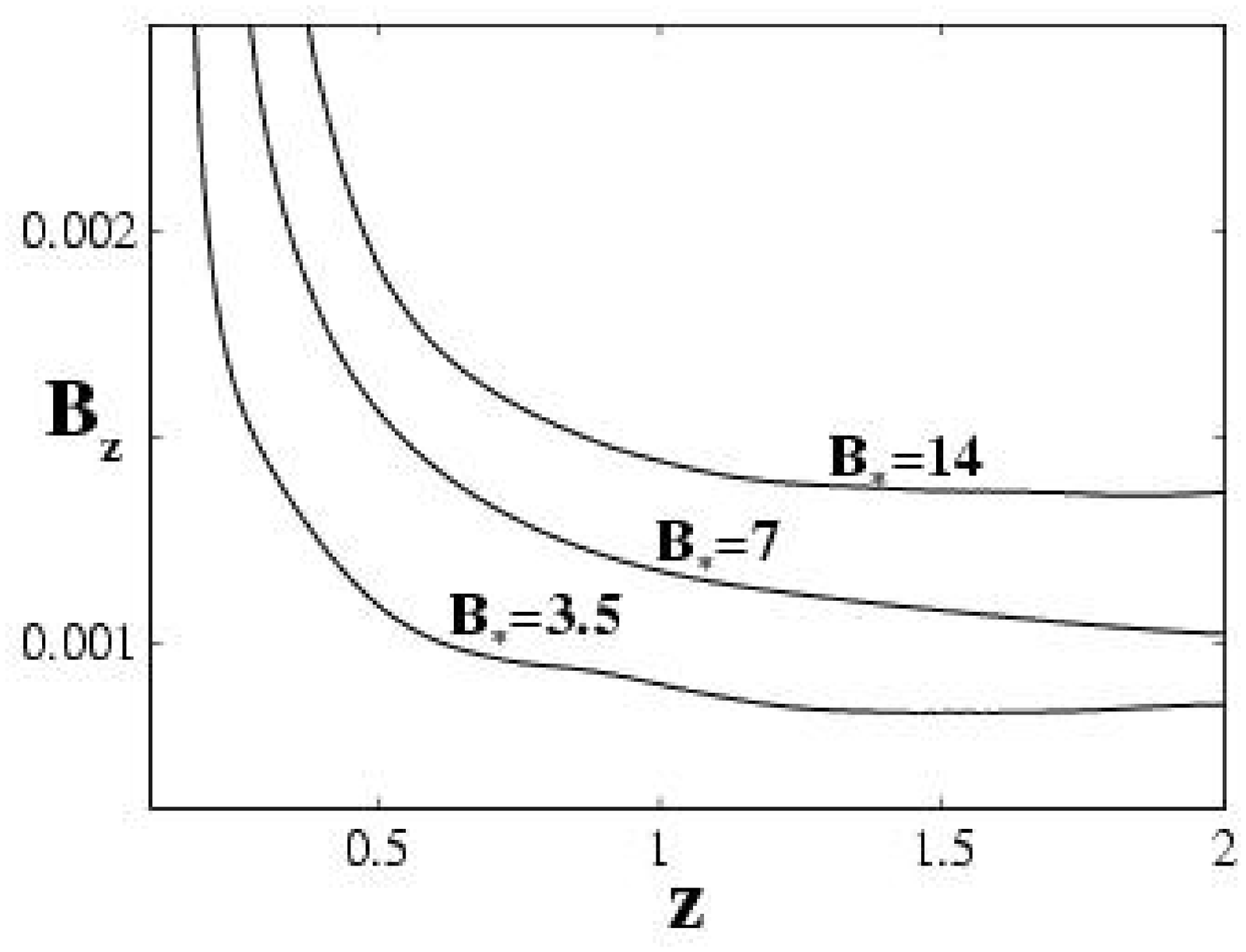,width=4.5in,height=3.5in}
\caption{ Variation  of the magnetic
field  along the tail at $r=0$ for $z \gtrsim 0.1$,
 ${\cal M}=3$ and different
values of the
magnetic field $B_*$
at the star's surface.}
\label{Figure 13}
\end{figure*}

\section {Accretion for $R_A \sim
R_{ \acc} $ and   ${\cal M}=3$}

   In this section we take
the Mach number to be
relatively small, ${\cal M}=3$, so that
the accretion radius
$R_{acc}$ is of the  order of magnitude
of the Alfv\'en radius $R_A$.

\subsection{Hydrodynamic
Simulations}

First, for reference, we did
hydrodynamic  simulations of the BHL
accretion to a {\it non-magnetized} star
for Mach number ${\cal M}=3$.
  We verified
that the nature of the flow is close
to that described by earlier
investigators of hydrodynamic BHL
accretion (e.g., Matsuda {\it et al.} 1991;
Ruffert 1994b).
   Namely, incoming matter forms a conical
shock wave around the star.
    Figure 2 shows
the main features of the flow at
a late time $t=
6.7t_0$ when the flow is stationary.
   The opening angle of the shock
wave at large distances from the star
relative to the $z-$axis is predicted to be
$\theta = {\rm arcsin}(1/{\cal M})$, which
is $\theta = 19.5^\circ$ for
${\cal M}=3$.
   Our simulations give
$\theta \approx 25^\circ$, which is larger
than  predicted.
   However, when we performed the
simulations in an enlarged region,
$R_{max}=2$, $Z_{min}=-2$, $Z_{max}=4$, on
a grid  $N_R\times N_Z =257\times 769$,
we obtained $\theta \approx 23^\circ$
which is close to the theoretical value
and similar to the value obtained by
Ruffert (1994b).

  We calculated the accretion rate $\dot M$ to
the numerical star and got a value
$\dot M\approx 0.5 \dot{M}_{BHL}$.
  We performed simulations using a
smaller numerical star
$R_*=0.02$ and got a slightly smaller value
$\dot{M} \approx 0.4 \dot M_{BHL}$.
   This behavior agrees with
Ruffert's results on the dependence of $\dot
M$ on  numerical star size
for the sizes used,
$R_*=0.25 R_{acc}$ and $R_*=0.1 R_{acc}$
(Ruffert 1994 a,b).
   This size dependence becomes
negligibly small for $R_* <0.1 R_{acc}$.
    In our simulations of accretion
to magnetized stars we  take the larger value
$R_*=0.05=0.25 R_{acc}$, because it gives
better resolution of
the magnetic field near the star.

    We compared simulations in
the region ($R_{max}=2 R_B$, $Z_{max}=4 R_B$)
with simulations in twice as smaller region.
    This gave only a
$\sim 5 \%$ decrease of accretion rate,
which means that our standard region
$R_{max}=2 R_B = 2$ is sufficiently large
to accumulate matter from the far distances, though
simulations in smaller regions will be also sufficient.
   Usually,  a low pressure
is arranged inside
the numerical star (e.g., Ruffert
1994 a,b).
   However, it is impossible to
perform MHD simulations with
very low pressure and density inside
 the numerical star.
    In our MHD simulations we
have $\rho=\rho_0=1$ inside the numerical
star.
   To test the influence
of this value, we performed
simulations with lower densities inside
numerical star,
$\rho_{acc}=10^{-2}\rho_0$ and
$10^{-3}\rho_0$
 We observed, that this
changed only slightly the accretion
rate (at the level  $< 5\%$).
   This is connected with the fact
that the matter density  which accumulates
around the star before accretion is much
larger than $\rho_0$, so that the difference
$\Delta \rho = \rho-\rho_0$ is about the
same for considered values of $\rho_0$.

\begin{figure*}[t]
\centering
\epsfig{file=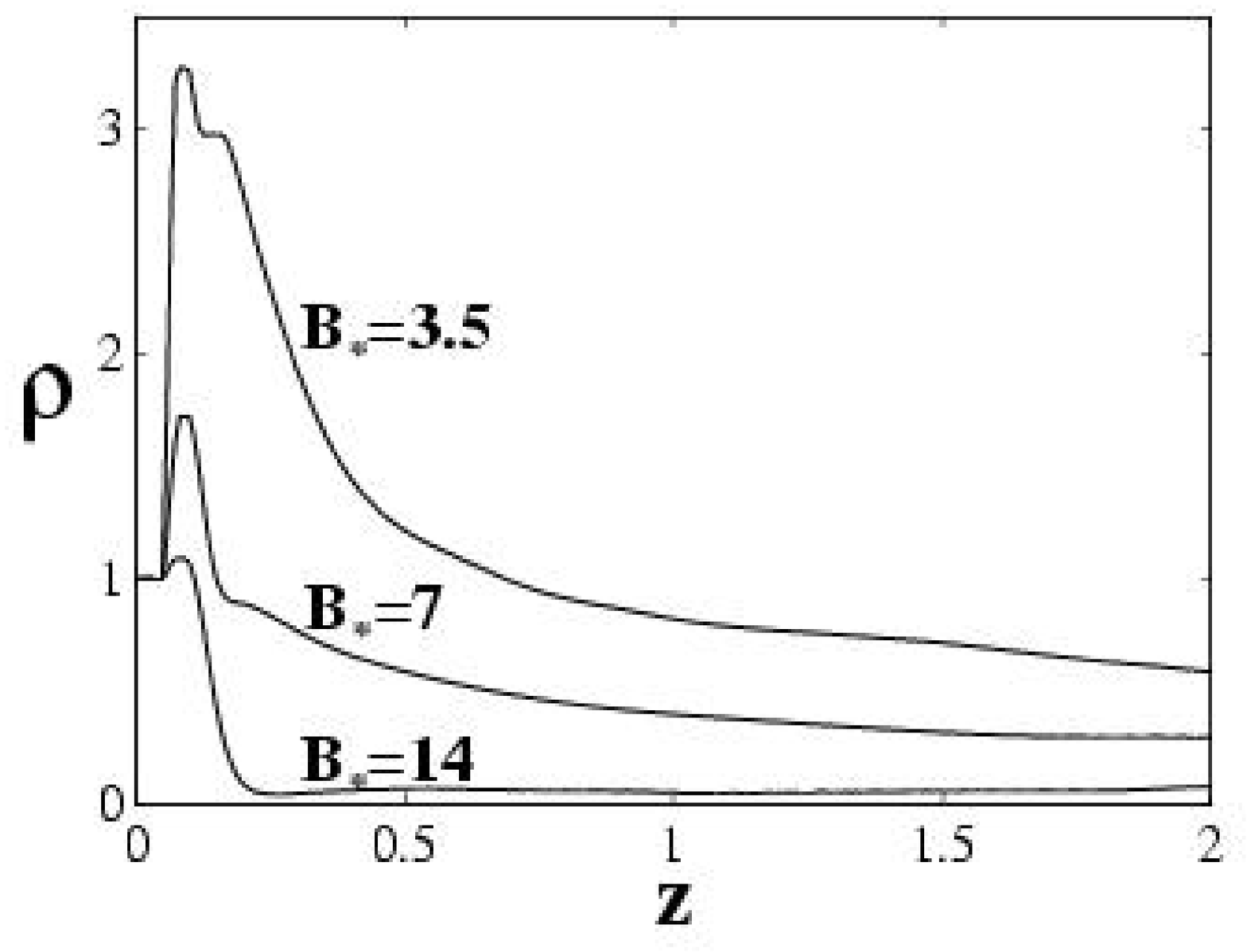,width=4.in,height=3.in}
\caption{ Axial distribution of density
in the tail for ${\cal M}=3$ and different
values of the magnetic field $B_*$.}
\label{Figure 14}
\end{figure*}

\subsection{Accretion to a Magnetized Star}

Next, we investigated propagation of a  {\it
magnetized} star through the ISM.
   Simulations were performed for
a number of  values
of the magnetic field
$B_*$.
   We  show
results for two cases: for a
relatively  weak magnetic field, $B_*=3.5$
(where $R_A < R_{acc}$);
and for a strong magnetic field, $B_*=14$ (where $R_A >
R_{acc}$).

   Figure 3 shows the main features
of the flow for a star with  $B_*=3.5$
at time $t= 5t_0$ when the flow is stationary.
    One can see that the magnetic
field of the star acts as
an obstacle for the flow and a
conical shock wave forms as in the hydrodynamic
case with similar angle $\theta$ as expected
since the Mach numbers are the same.
   Magnetic field lines (with
flux values the same as in Figure 1)
are slightly stretched by the flow, but they
remain closed.
    Figure 4 shows the inner region
of the flow in greater detail.
   The bold line represents
the Alfv\'en surface, where the matter
energy-density
$\rho(\varepsilon+ v^2\!/\!2)$ is equal to
the magnetic energy-density
$B^2/(8\pi)$.
   The radius of Alfv\'en surface
in $z-$ direction downstream at $r=0$ is
$R_A \approx 0.1$, and in  $r-$ direction at
$z=0$  is $R_A \approx 0.14$ which is
smaller than accretion radius
$R_{acc}\approx 0.2$.
   Thus, some
gravitational focusing is expected and indeed
we observe density enhancement around
the star.
   Figures 5 and 6 show the distribution
of magnetic flux with the lower  limit
${\rm log}_{10} \Psi_{min} = -6$
compared to that
shown in Figures 3 and 4 where ${\rm log}_{10}
\Psi_{min} \approx -5.3$.
    Thus the apparent
truncation of the magnetosphere in
Figures 3 and 4 was connected with choice of
minimum plotted magnetic flux.
     Streamlines
of matter flow $\rho{\bf v}$
shown in Figures 5 and 6 reveal that
matter from radii $r < 0.1 R_{acc}$,
accretes to the star, while the rest of
matter flies away.
     Compared with the
non-magnetized case, the magnetic field
acts as  an
obstacle for the flow, and most of the
inflowing matter is kept away from the star.

    The matter density is
strongly enhanced in the
shock wave, but gradually decreases
as it approaches  the
surface of the star where it accretes (Figure
7a).
     Behind the star (for $0<z<0.4$)
 there is also an accumulation of matter
connected with gravitational focusing by
a star.
    Note, that in the case of hydrodynamic
accretion, the density jump in front
of  the star
(at $z<0$) is much smaller, while
behind the
star (at $z>0$) it is much larger.
   The velocity
$v_z$ (panel b) decreases sharply in the
shock wave to small subsonic values, but
later increases again in the polar column.
   Behind the star the velocity
is negative in the
small region $0.05 < z < 0.1$,
where accretion
occurs.
   The density and velocity jumps in
front of the star do not satisfy the
standard Rankine-Hugoniot conditions,
because the shock
wave is ``attached'' to the magnetosphere.
   Matter cannot move freely after passage
through the
shock wave, and extra matter accumulation
occurs in the shock.

  From panels (a) and (b) of Figure 7
it is clear that
the  rate of accretion is smaller in
the case of a magnetized star compared to a
non-magnetized star.
   We observed, that the
accretion rate to a magnetized star for
$B_*=3.5$ is about $3$ times smaller than
that to a non-magnetized star.
  The variations of the
of energy-densities along and across the tail
(Figure 7c, d) shows that magnetic
energy-density dominates only in a small
region around the star.

\begin{figure*}[t]
\centering
\epsfig{file=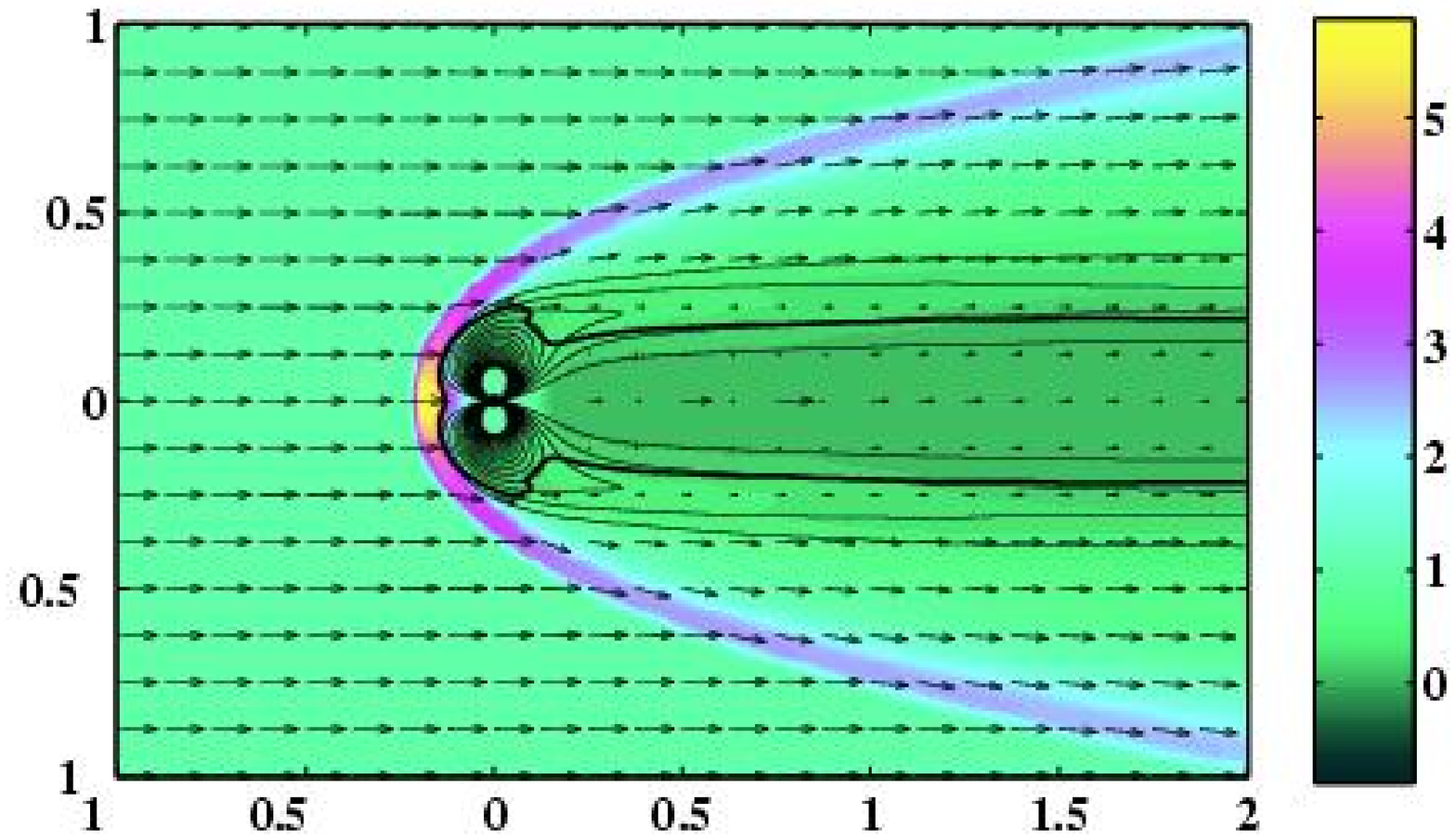,width=4.9in,height=3.in}
\caption{ Results of simulations for ${\cal
M}=10$ and $B_*=14$.}
\label{Figure 15}
\end{figure*}

\begin{figure*}[b]
\centering
\epsfig{file=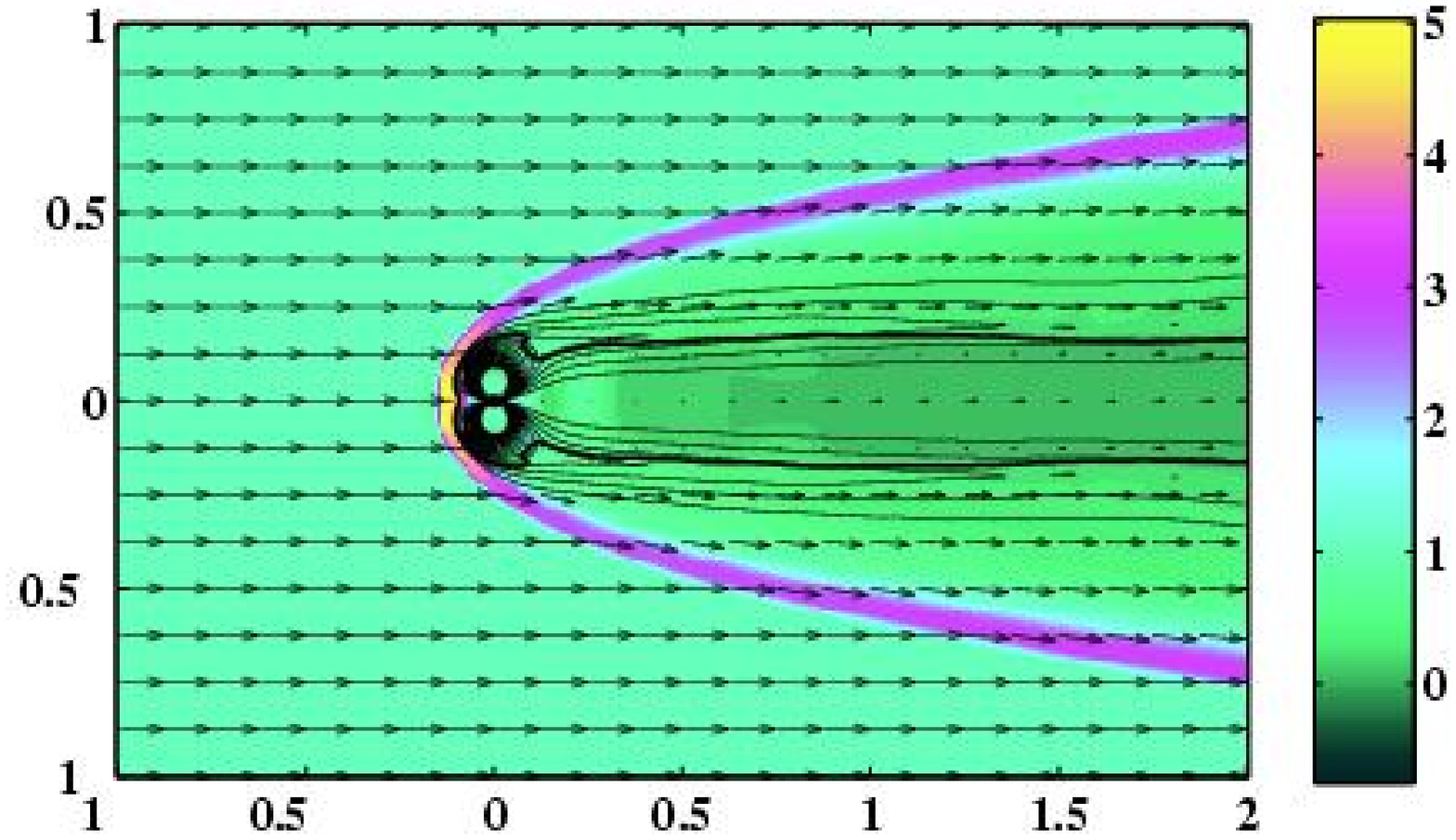,width=4.9in,height=3.in}
\caption{ Results of simulations for ${\cal
M}=30$ and $B_*=14$.}
\label{Figure 16}
\end{figure*}

\begin{figure*}[t]
\centering
\epsfig{file=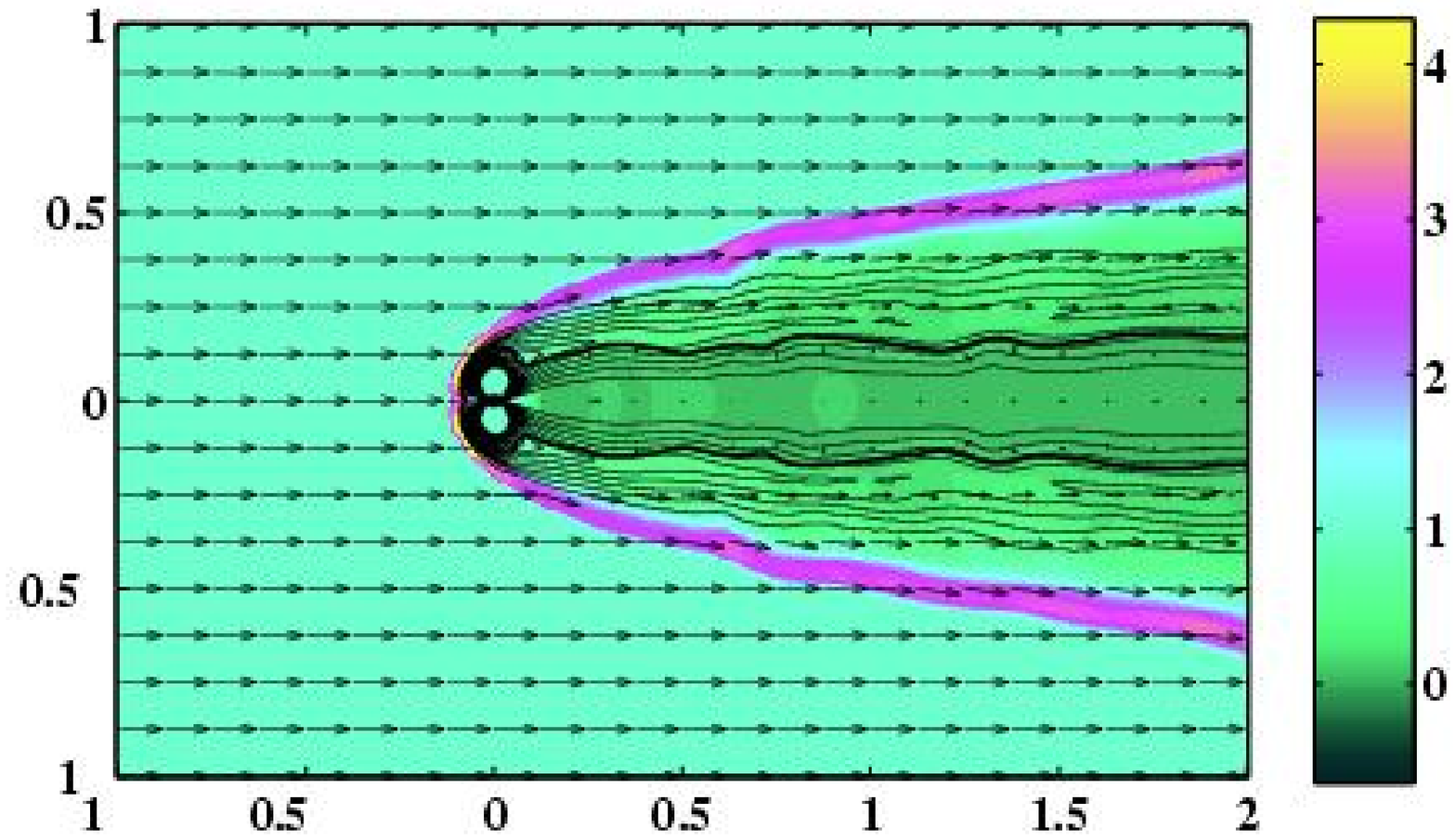,width=4.9in,height=3.in}
\caption{ Results of simulations for ${\cal
M}=50$ and $B_*=14$.}
\label{Figure 17}
\end{figure*}

   In case of a stronger magnetic field,
$B_*=14$, larger magnetic flux
is stretched downwind (see Figure 8).
   Gravitational
focusing is still important and density
enhancement is observed around the star, but
it is much smaller, than in case of weaker
magnetic field $B_*=3.5$.
    Now, the
Alfv\'en surface has elongated structure
and extends all along the $z-$axis, so that
the magnetic energy-density predominates in
the tail (see Figure 8).

    The magnetosphere around the star 
is larger than in the
case $B_*=3.5$, and the Alfv\'en
radius in $r$ direction is $R_A\approx
0.26$, which is larger than accretion radius
$R_{acc}=0.2$ (Figure 9).
   Now, all incoming
matter goes around the magnetosphere and
flies away.
   Streamlines of the flow (Figure 10) show,
that no matter goes from the
front and accretes to the back side
of the star.
   A small flux
of matter coming from $r << R_{acc}$,
accretes directly to the upwind pole of
the star.

\begin{figure*}[t]
\centering
\epsfig{file=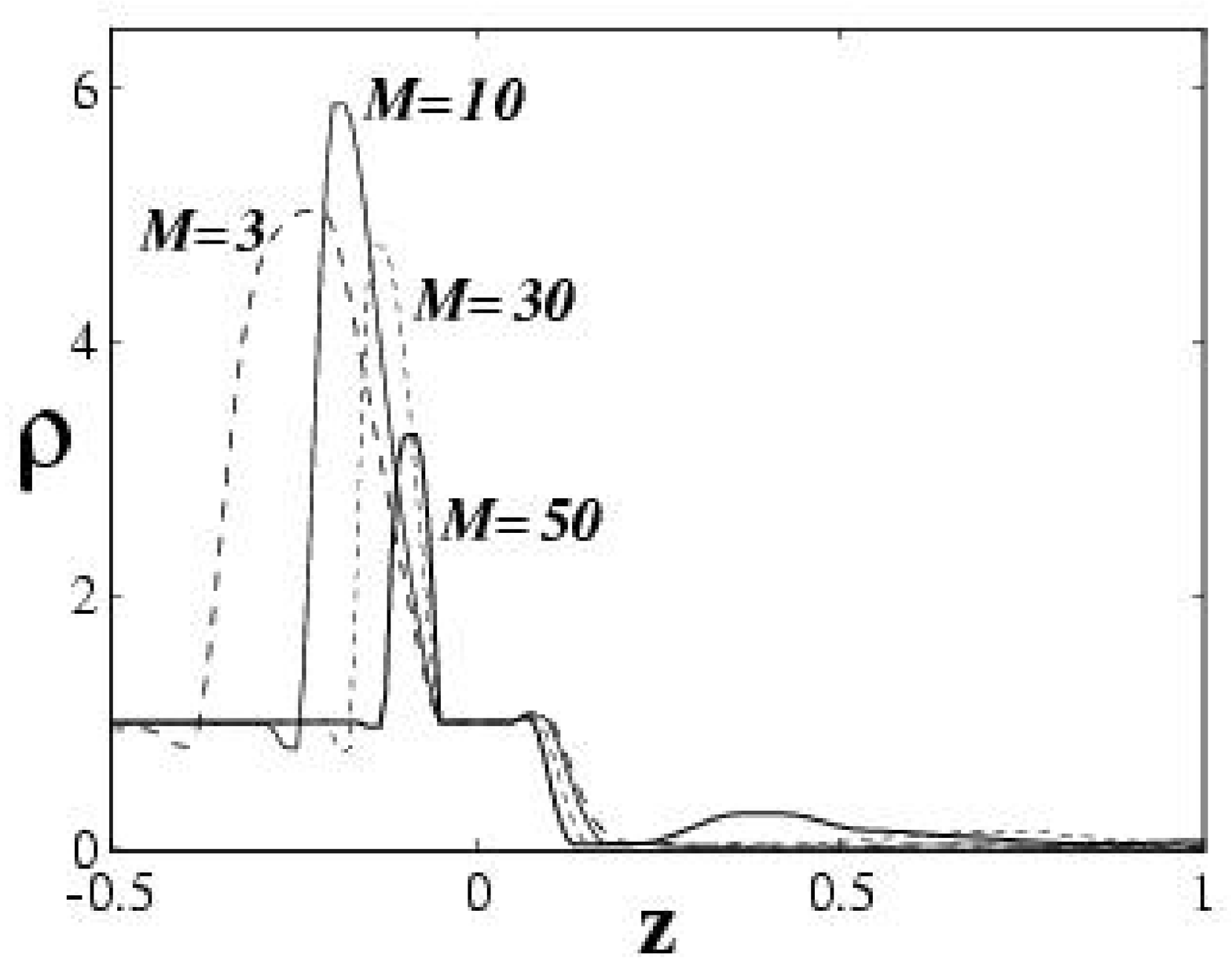,width=4.in,height=3.in}
\caption{Axial density variations at
different  Mach numbers and $B_*=14$.}
\label{Figure 18}
\end{figure*}

Figure 11a shows, that at $B_*=14$, compared to
$B_*=3.5$, magnetic
energy-density predominates in the tail in the
region of equatorial plane. Figure 11b and also
Figure 8 and 9, show that in $r-$ direction
magnetic energy-density dominates in the tube with
radius $r\approx 0.11$.

     We performed additional  simulations for
magnetic fields strengths
$B_* = 2$, $7$, and  $11$, and derived the dependence of
the accretion rate on
the magnetic field strength $B_*$ for
all cases.
    We observed that accretion rate
strongly decreases with increasing magnetic
field (see Figure 12) as $\dot M \sim
B_*^{-1.3\pm 0.05}$.
     This dependence reflects the fact that
a stronger magnetic field of the star deflects
the incoming ISM flow more efficiently than
weaker magnetic field.

     Figure 13 shows axial variation of $B_z$
for different values of $B_*$.
     In all cases the magnetic field
decreases very gradually with $z$:
$B_z\sim z^{-0.15}$. The decrease is
partially connected with gradual radial
expansion of the magnetosphere, and
partially with the reconnection of magnetic
field lines in the tail.
   In the actual flow, the magnetic
diffusivity is expected to be
much smaller than that in the
code.
   This acts to decrease the reconnection rate
and increase the length of the tail.
    Note that the tail of Earth's magnetosphere
extends to more than a hundred of Earth
radii (e.g., Nishida {\it et al.} 1998).
   The value
of the field in the tail is larger for larger
values of  $B_*$.
    Even for
$B_*=3.5$ the magnetic field stretches
a long distance downwind from the star.
    The Alfv\'en surface
in this case is small not only because the
magnetic field is weak, but also because
matter energy-density is high.
   At magnetic field strengths
$B_* < 2-3$, however, stretching of magnetic
field to the tail becomes suppressed.

\begin{figure*}[t]
\centering
\epsfig{file=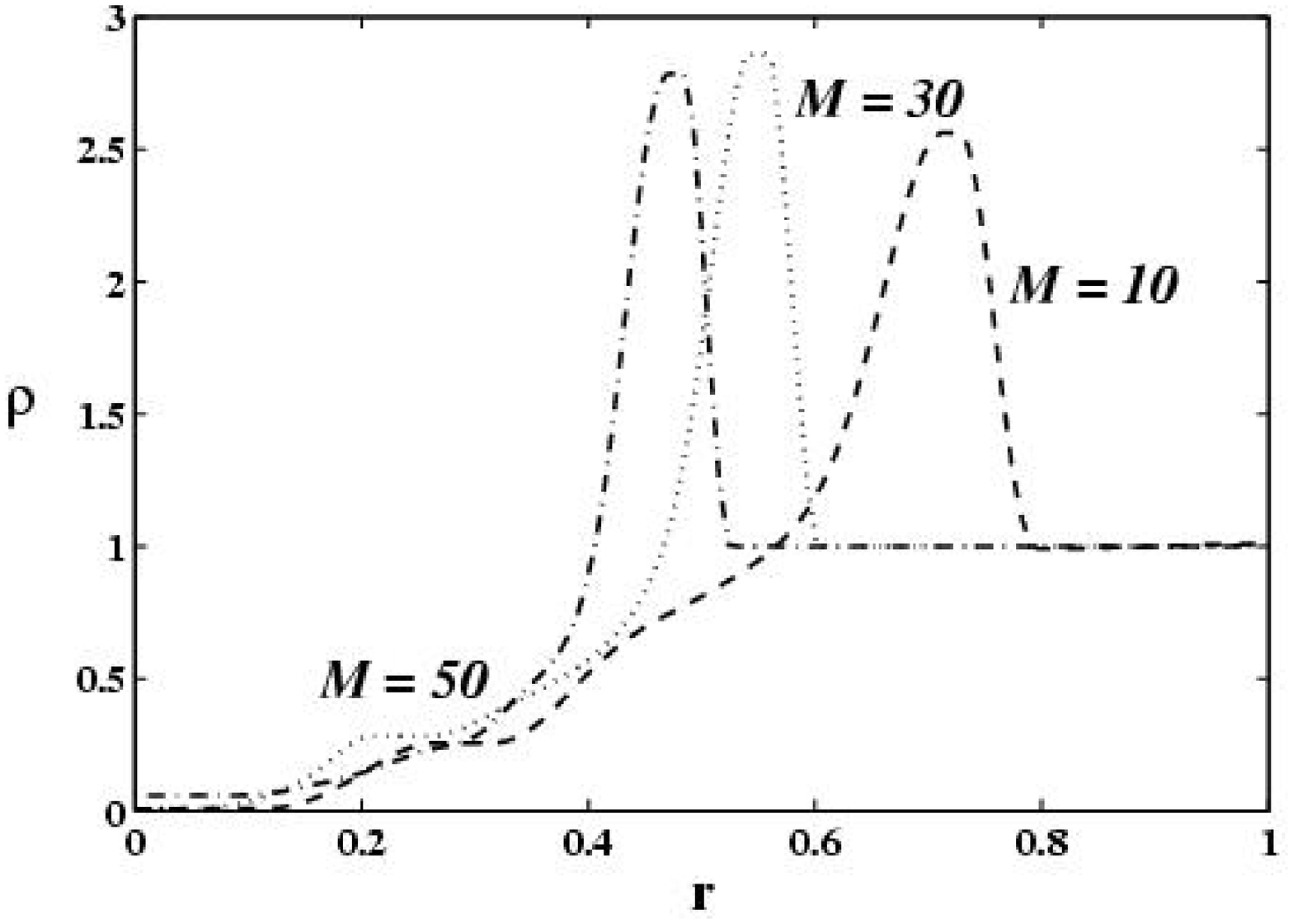,width=4.in,height=3.in}
\caption{ Radial density variations at
$z=1$ for
different Mach numbers and $B_*=14$.}
\label{Figure 19}
\end{figure*}

   For $B_*>7$ ($R_A \gtrsim R_{acc}$), the
density in the magnetotail is lower than
that in the incoming flow
$\rho_0$ and it decreases at higher $B_*$
 (see Figure 14).
    Thus, one can expect
hollow magnetic tails in the case of strongly
magnetized stars. This is connected with the
fact, that magnetosphere is an obstacle for
the flow and the tail represents the
rarefaction region which usually forms
behind an obstacle in a supersonic
flow (e.g. Landau \& Lifshitz 1960).
    Furthermore, external matter
penetrates only slowly across the  magnetotail,
because  the magnetic diffusion
time-scale across the tail is
long compared with the transit time of
the matter in the $z-$direction.

\section{``Magnetic Plow" Regime ($R_A >>
R_{\acc}$)}

   In this section we investigate interaction
of magnetosphere with the ISM in ``magnetic
plow" regime, where the
 Alfv\'en radius $R_A$ is much larger than
accretion radius
$R_{acc}$.
   In this limit gravitational focusing is
unimportant and there is only direct
interaction of the ISM with magnetosphere of
the star.
   For Mach numbers larger than about
${\cal M} = 3$ (for our set of parameters
$B_*$), the flow is in the ``magnetic plow" regime.
     In this section we investigate
 properties of magnetotails at different
Mach numbers ($\S 5.1$) and different magnetic
diffusivities ($\S 5.2$).

\begin{figure*}[t]
\centering
\epsfig{file=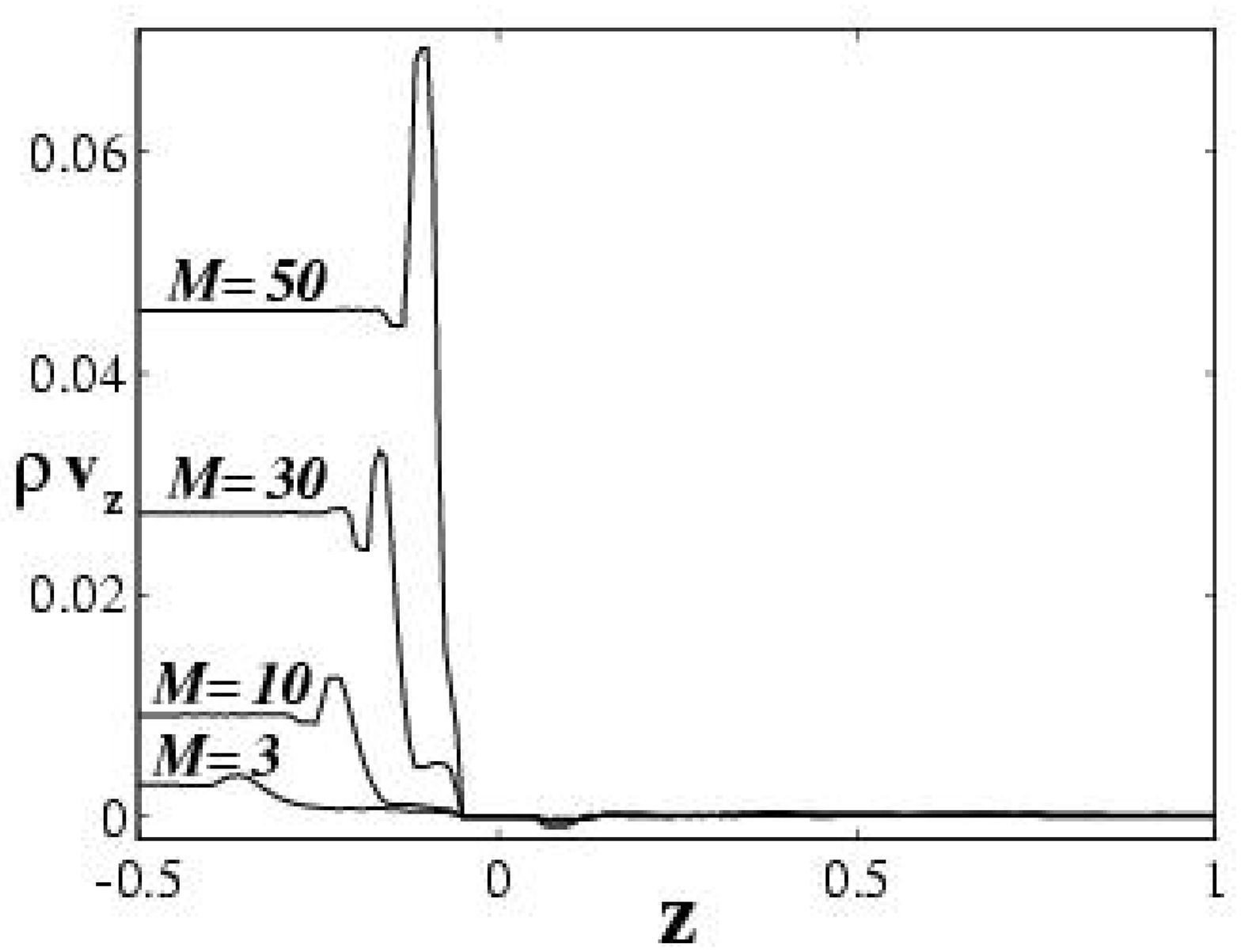,width=4.in,height=3.in}
\caption{Axial  variation  of the matter flux
$\rho v_z$
 at different Mach numbers ${\cal M}$ and $B_*=14$.}
\label{Figure 20}
\end{figure*}

\subsection{Investigation of Magnetotails at
Different Mach Numbers ${\cal M}$}

   In this subsection we
fix the magnetic field to be
$B_*=14$ and the diffusivity
$\tilde\eta_m=10^{-6}$ and investigate
flows at Mach numbers
${\cal M}=10$, ${\cal M}=30$ and  ${\cal
M}=50$.
    We observed, that at high Mach
numbers ${\cal M}$, the sharp density
enhancement is observed in the shock cone,
while the rest of the tail have low density
(see Figures 15, 16 \& 17).
     At very high Mach number ${\cal M}=50$,
some kind of instability appears in the tail
which determines its wavy behavior (Figure 17).
    This instability may be connected with high
velocity gradient across the tail.
     The Alfv\'en radius in the  $r$ direction
and in the
upwind $z$ direction decreases at larger
$\cal M$ (see also equation 8), because the
flow strips deeper layers of magnetosphere.
This also leads to higher magnetic field in
the tail.
        Reconnection is
observed as in case of lower Mach numbers.
    However, the reconnection region is
further downwind from the star at higher $\cal M$.

   The axial density variations for the three
cases are shown at Figure 18.
    The case with low
Mach number ${\cal M}=3$ is included for
reference.
    One can see that in case ${\cal M}=10$
the density in front of the
star increases to $\rho_{front}=(5-6)
\rho_0$ and then decreases sharply closer to
the surface of numerical star.
   At higher Mach numbers,
the density peak is lower.
      The density behind
the star, in the tail, is small
$\rho_{tail}\sim (10^{-1} -10^{-2}) \rho_0$.
    The density variation across the tail at
$z=1$ is shown at Figure 19.
    It shows that an
essential part of the tail is hollow.
   The matter
flux $\rho {\bf v}$ is much higher for higher Mach
numbers (Figure 20), owing to higher
velocities.
    The instability observed
at ${\cal M}=50$ may be the
Kelvin-Helmholtz instability connected
with the large gradient in the flow velocity.

\begin{figure*}[t]
\centering
\epsfig{file=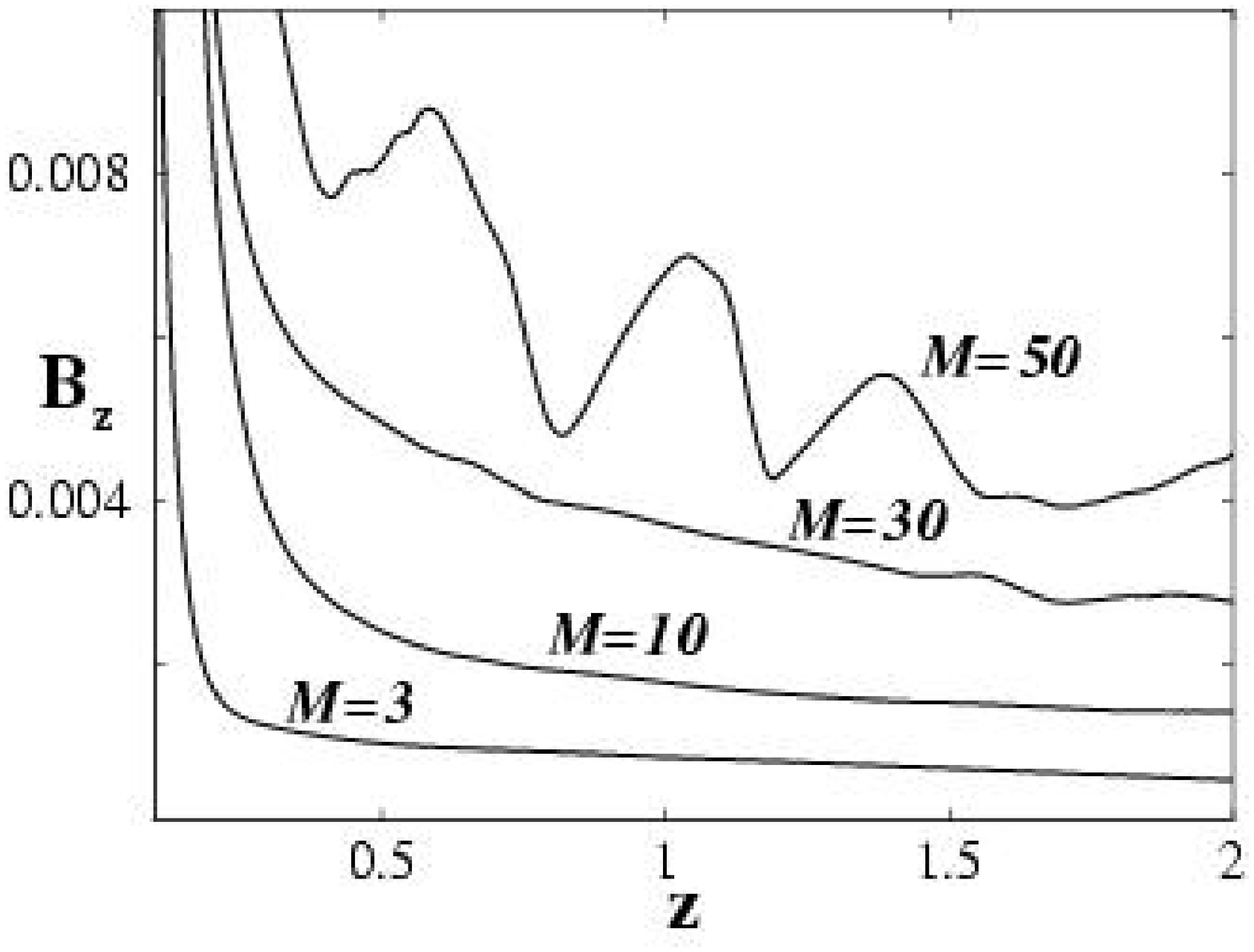,width=4.in,height=3.in}
\caption{ Variation of $B_z$ along the tail at $r=0$
for $B_*=14$ for different Mach numbers.}
\label{Figure 21}
\end{figure*}

   The axial magnetic field  decreases slowly
with distance behind the star,
$B_z\sim z^{-0.2}$ (Figure 21).
 Thus, long tails form as in the
case  ${\cal M}=3$.
   The magnetic field in the tail is
larger at larger Mach numbers.

\begin{figure*}[t]
\centering
\epsfig{file=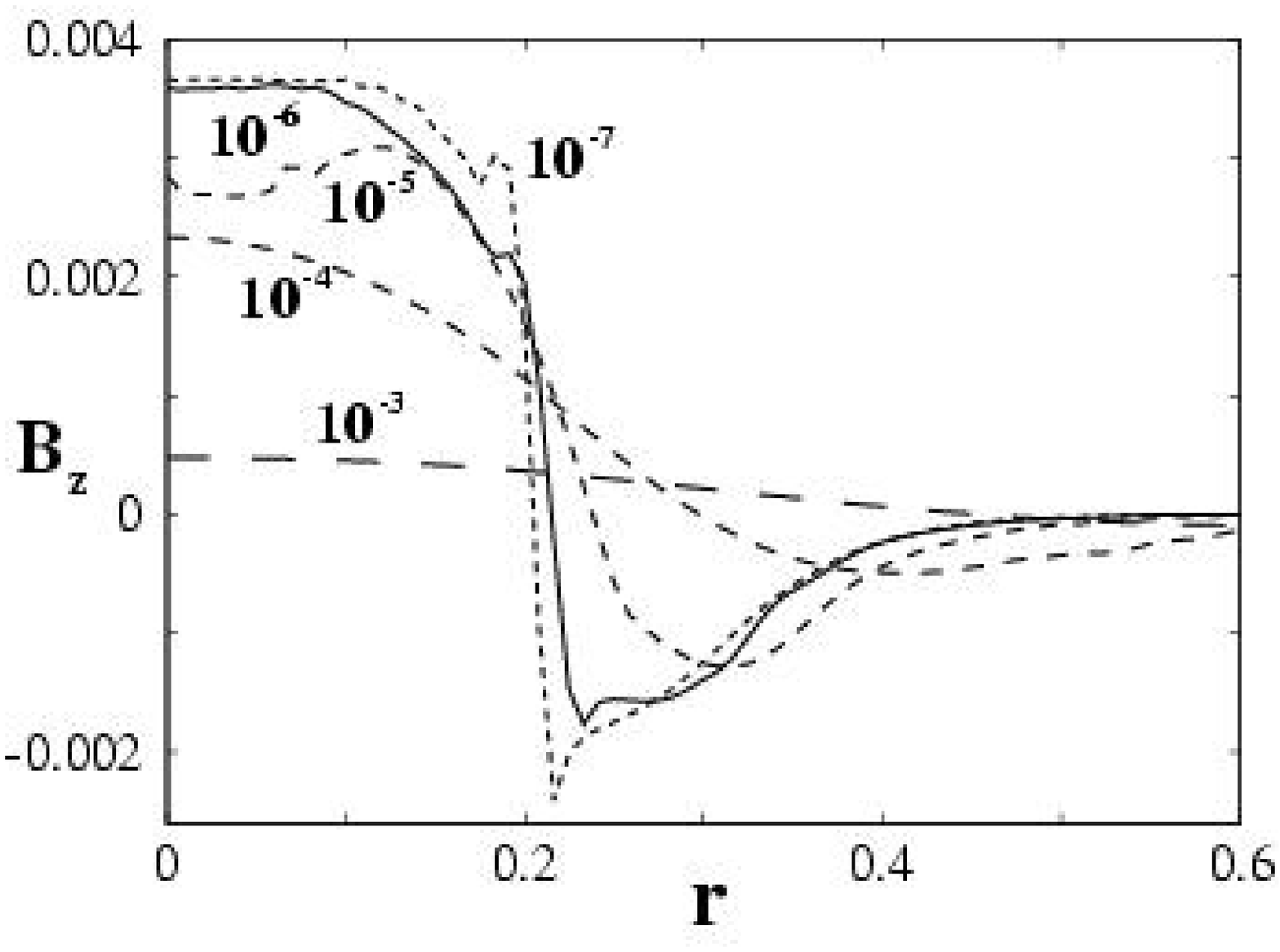,width=4.in,height=3.in}
\caption{Variation  of the magnetic field
$B_z$ across the tail at $z=1$ for different
values of the magnetic diffusivity
$\tilde\eta_m$ for the case $B_*=14$ and
${\cal M}=30$.}
\label{Figure 22}
\end{figure*}

\begin{figure*}[t]
\centering
\epsfig{file=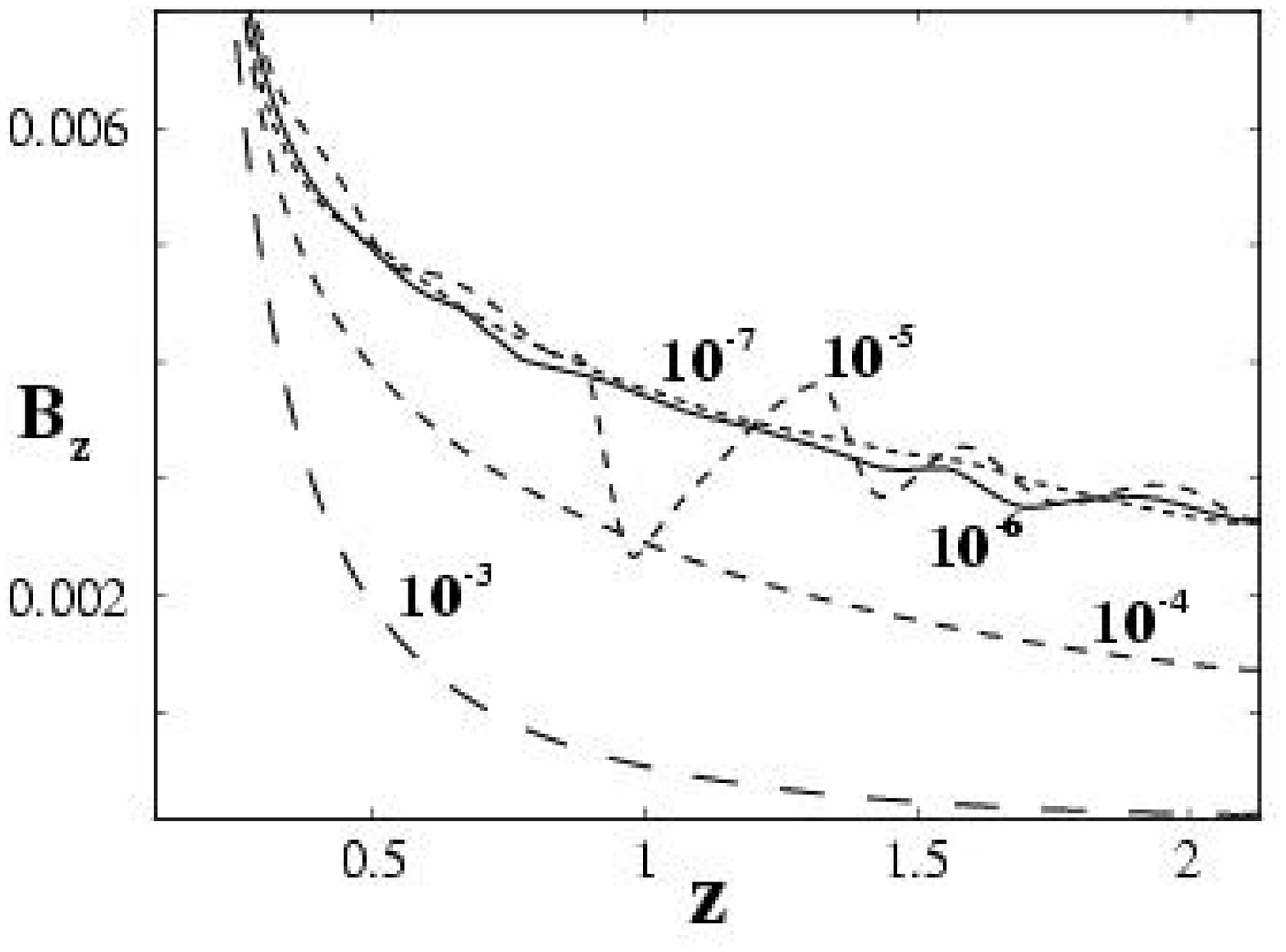,width=4.in,height=3.in}
\caption{Variation  of the magnetic field
$B_z$ along the tail at $r=0$
for different values of the
magnetic diffusivity
$\tilde\eta_m$ for the case $B_*=14$ and
${\cal M}=30$.}
\label{Figure 23}
\end{figure*}

\subsection{Dependence of the Flow on
Magnetic Diffusivity}

    The processes  of accretion and
reconnection of the magnetic field
depend on the magnetic
diffusivity $\tilde\eta_m$.
    The fact that our code
explicitly includes $\tilde \eta_m$
allows us to investigate the
dependence of the flows on the
magnitude of this quantity.
   This is in contrast with
ideal MHD codes where the
magnetic
diffusivity unavoidably
arises from the finite
numerical grid.
    To study the dependence
on $\tilde \eta_m$, we fixed the magnetic
field, $B_*=14$, and the Mach number, ${\cal
M}=30$.
    We  made simulation runs
for a range of values between
 $\tilde\eta_m = 10^{-3}$ and
$10^{-8}$.

    We observed
that at lower magnetic
diffusivity, the magnetic tail
(the Alfv\'en surface) is wider in the $r$
direction.
    Figure 22 shows the  variation of $B_z$
across the tail at $z=1$.
   One can see that
at small $\tilde\eta_m=10^{-6} - 10^{-7}$,
regions with oppositely directed magnetic
field are very close to each other, but do not
reconnect.
   On the other hand, at large
$\tilde\eta_m=10^{-3} - 10^{-4}$, the magnetic
field is much smaller,
because it annihilates rapidly with
distance behind the star.
    Furthermore, note that at large $\tilde\eta_m$,
matter is partially decoupled from magnetic
field and stretching of magnetic field is
less efficient.
   Figure 23 shows the dependence of
axial distribution of $B_z$ on
$\tilde\eta_m$.
   One can see that at
$\tilde\eta_m > 10^{-5}$,  the magnetic field
decreases with $z$ very rapidly.
   Note, that at
$\tilde\eta_m > 3\times 10^{-7}$, numerical
diffusivity predominates, and the calculated
flows depend
only weakly depends on
$\tilde\eta_m$.

    The observed behavior is
determined by the magnetic
Reynolds number,
$$
     Re_m \equiv {R~v \over\eta_m} =
{\tilde R ~ \tilde
v \over \tilde{\eta}_m}~,
\eqno(16)
$$
where tilde-quantities are our
dimensionless variables.
   For example, for ${\cal M}=3$ and
$B_*=3.5$, in the upwind region of the flow,
$Re_m\approx
400$ and most of
the matter goes around the dipole
and flies away or accretes to the downwind
pole.
   Matter which goes to the downwind
pole has  smaller velocity and hence
smaller $Re_m$.
   Also, when matter diffuses
across the tail in the $r$ direction owing to
gravitational force it has $v_r << v_z$, and
$Re_m \lesssim 1$.
   However, the timescale of the flow
in the $z-$ direction is
much less than that in
$r$ direction, so that most of the matter flies
away.
   The main conclusion of this subsection is
that the magnetotails lengthen as
the diffusivity decreases.
    Comparison with the Earth's
magnetosphere (e.g., Nishida {\it et al.} 1998),
shows that the actual diffusivity may
be smaller than the smallest
values used in our simulations.

\section{Observational Consequences}

The question arises, is it possible to
observe either a bow shocks or the
elongated magnetotails of
magnetized old neutron stars or
magnetars?
     In this section we  estimate the
powers released
and  other possible observational
features of these objects.

\subsection{Reconnection in the Tail}

Our  simulations show that the
magnetic field in the tail reconnects.
   This phenomenon may
lead to acceleration of particles and
possible flares in the tail.
   The total magnetic energy
stored in the tail can be estimated as
$$
E_{tot} \approx {1 \over 8\pi}\int\limits_0^S
dz~\pi [R(z)]^2 ~
[B(z)]^2 ,
\eqno(17)
$$ where $S$ is the length of the tail,
$R(z)$ is the radius of the tail at  $z$.
   The total magnetic flux in, say, the
$+z$ direction along the tail,
$$
\Phi_{mag}\approx B(z) \pi [R(z)]^2 \approx B_A \pi
R_A^2~,
\eqno(18)
$$
is constant in the absence of reconnection.
   Therefore, if the tail cross-section
expands with
distance $z$, then the magnetic
field decreases as
$B(z)=B_A (R_A/R(z))^2$.
   The values $R_A$
and $B_A$ we derived earlier [see equations (8)
and (9)].
   We observed that
at high Mach numbers, the magnetotail
expands in the $r$
direction very gradually.
   To estimate the
total magnetic energy in a
tail of length $S$, we suppose that the tail
does not expand, ($R_{tail}\approx R_A$) thus
$$
E_{tot}\approx {1\over 8 \pi}{B_A^2} \pi R_A^2 S
\sim
10^{27} B_{12} n^{1/2} v_{200}~
S_{100} ~{\rm erg}~,
\eqno(19)
$$
where $S_{100}=S/(100 R_A)$.

  Two main physical processes  determine
the length of the magnetotail.
   The first  is the
stretching of magnetic field lines by the
incoming matter flow.
   This mechanism operates on the
dynamical time-scale,
$$
t_{dyn}=\frac{S}{v} \sim 10^6
B_{12}^{1/3} n^{-1/6} v_{200}^{-4/3}
S_{100}  ~{\rm s}.
\eqno(20)
$$
   The stretched tail magnetic field has
regions of opposite polarity so that the total
magnetic flux in the $z$ direction is zero.
      In the axisymmetric
case studied here, a cylindrical neutral
layer  forms.
    Magnetic
field reconnection/annihilation may occur
all along this layer.
    The length $S$ of the tail is
determined by the competition between
stretching and reconnection of the magnetic
field.
    A nominal time-scale for reconnection
across the tail is $t_{dif}= R^2/\eta_m$.
    In view of equation (16),
$t_{dif}/t_{dyn} = Re_m (R/S)$.
   A balance between the stretching and
diffusion implies that this ration is
of order unity.
    With $t_{dif}\approx
t_{dyn}$, the average power
released by reconnection is
$$
\dot{E}_{rec}\approx \frac{E_{tot}}{t_{dyn}} \sim
10^{21} {B_{12}}^{2/3} n^{2/3}
v_{200}^{7/3}~{\rm \frac{erg}{s}}~.
\eqno(21)
$$

   Next, we estimate the
power released in an individual ``flare,''
which is termed a ``substorm'' in
the case of the Earth's magnetotail.
If such a flare occurs in
a cylinderical   volume
$\sim \pi {R_A}^3$, then the  energy
released is
$$
E_{rec} \sim \frac{{B_A}^2}{8\pi} \pi R_A^3
\sim
10^{25} B_{12} n^{1/2} v_{200} ~
{\rm erg}~.
\eqno(22)
$$
The power of the flare, $\dot
E_{rec}=E_{rec}/t_A$, depends on the
reconnection time-scale
$t_{rec}=R_A/v_A$,
where $v_A=B_A/\sqrt{4\pi\rho}$ is Alfv\'en
speed.
    The Alfv\'en speed is a function of
density $\rho$ which is uncertain.
     Our
simulations  show that the density in
the tail is much lower than the density of
incoming ISM.
    It decreases as the
magnetic field $B_*$
increases (see Figure 14).
    We have not been able to do
simulations for very strong magnetic
fields such as those of
magnetars.
    However,
the uncertainty in $n_{tail}$ can
be handled by looking at the
extreme cases:  (1) a relatively high density
tail where $n_{tail}=1/{\rm cm}^{3}$; and
(2) a very low density tail where the Alfv\'en
velocity approaches the speed of light
$v_A \lesssim c$.
This density is
$ n_{tail}\approx 4.4\times 10^{-7} n
v_{200}^2 ~{\rm cm^{-3}}.
$

    For the case of a high matter density
in the tail, we get
$v_{rec}\approx v_A$ and
$$
t_{rec}\sim10^4 B_{12}^{1/3}
n^{-1/6} v_{200}^{-4/3}  ~{\rm s}~,
\eqno(23)
$$
and
$$
\dot E_{rec} \sim 10^{21}
B_{12}^{2/3} n^{2/3} v_{200}^{7/3} ~{\rm
erg/s}~.
\eqno(24)
$$
  Note, that this power coincides with
our estimate [equation (15)] based
on the dynamical time-scale.

In the case of low density tail,
we find
$$
t_{rec} \sim 7.4 B_{12}^{1/3} n^{-1/6}
v_{200}^{-1/3} ~{\rm s}~,
\eqno(25)
$$
and the power
$$
\dot E_{rec} \sim 1.6\times 10^{24}
B_{12}^{2/3} n^{2/3} v_{200}^{4/3}~ {\rm
erg/s}~.
\eqno(26)
$$
Thus, the power released in individual
flares is small even in the
case of the fastest
reconnection rate.
   The radiation spectrum
of released energy is
unknown.
    In view of the weak
magnetic fields in the tail,
$B_{tail}\sim 10^{-4} - 10^{-6}$G,
and the possible very low densities,
the energy may go into accelerating
electrons which then radiate
in the radio band.

\subsection{Bow Shock Radiation}

   Part of the power output of
a high Mach number magnetized
star is released in the bow
show wave where
the heated ISM behind
the shock radiates.
    The total power released at
the front part of the shock,
$r \lesssim R_A$, is
$$
\dot E_{shock} \approx {\pi \over 2} R_A^2
\rho v^3 \sim 10^{21}
n^{2/ 3} v_{200}^{7/ 3}
B_{12}^{2/ 3}~ {\rm erg/s}~.
\eqno(27)
$$
This power is comparable to
the steady power released
by reconnection in the magnetotail.
  The post shock temperature
is $T\approx m_p v^2/{3k}\approx 1.6\times
10^6 v_{200}~{\bf K}$ which corresponds to X-ray
band.
   The ISM particles excite hydrogen
atoms which re-radiate in the optical and UV
bands. Thus, one expects radiation
from the shock wave from the optical to X-ray
bands.

\subsection{Astrophysical Example}

    In this paragraph we give the connection
between the simulation parameters and the
astrophysical quantities.
  The density of the ISM is taken as
$n_\infty=n=1~{\rm cm}^{-3}$
and the sound speed as
$c_{s\infty}$.
     Then, from equation (14)
we obtain the reference
magnetic field
 $B_0 \approx 0.015 n^{1/2}
(c_{s\infty}/30{\rm km/s}) \beta_{-6}^{-1/2}~{\rm G}$,
where $\beta_{-6} \equiv \beta/10^{-6}$.
   For example, if the
dimensionless field is $B_*$, then the
actual magnetic field is
 $B \approx 0.015 B_* n^{1/2} (c_{s\infty}/30{\rm km/s})
\beta_{-6}^{-1/2}~{\rm G}
$ at the radius
$R=0.25 R_* \approx 0.0125 R_B=2.6\times 10^{11}
(c_{s\infty}/30{\rm km/s})^{-2}~{\rm cm}$, which correspond to
an external region of the actual magnetosphere.
    We can extrapolate this
field to smaller radii to get
the magnetic field at the surface
of the star with radius
$R_s=10~{\rm km}$:
$B_s \approx
2.6\times 10^{14} B_* n^{1/2} (30{\rm km/s}/c_{s\infty})^{5}
\beta_{-6}^{-1/2}~{\rm G}$.

\subsection{Comparison with Earth's
Magnetosphere}

   There are  similarities and differences
between the supersonic solar wind
interaction with the Earth's
magnetosphere and the interaction
of the ISM with pulsars.
   The magnetization of the solar
wind is important for the interaction
with the Earth's magnetic field.
   Although not included in the present
study, the magnetization of the ISM
may also be important for the interactions
with the neutron star magnetosphere.
   In contrast with the solar wind -Earth interaction, the
Mach numbers of pulsars vary from
${\cal M}\sim 1$ to ${\cal M}\sim 150$
for the fastest pulsars (Cordes \& Chernoff
1998).
    The orientation angles
of magnetic axes   $\theta$ relative to the
propagation
direction vary from $\theta=0^\circ$
to $\theta=90^\circ$.
    If the  high velocities of
some pulsars
are connected with initial magnetic
or neutrino kicks
(Lai, Chernoff \& Cordes 2001), then
one may expect this angle to be closer to
$\theta\approx 0$, similar to that
considered in this paper.

\subsection{Observational Consequences of
Long Hollow Tails}

   The discussed simulations
have shown that a long
 hollow, low-density magnetotail
forms behind a high Mach number
magnetized star.
   This fact, and the fact that in the
magnetic field lines are highly
stretched in the tail,
leads to possibility that
particles accelerated near the star can
preferentially propagate along the tail.
  This effect may also be important during
pulsar stage.
   A pulsar generates a relativistic
wind consisting of magnetic field and
relativistic particles (Goldreich \& Julian
1969).
   The standoff distance of the shock
wave is determined by the total power
generated near the light cylinder (e.g.,
Cordes {\it et al.} 1993).
   Significant part of
energy may be in magnetic field.
  Expanded
magnetospheres of pulsars interact with the
ISM forming elongated structures but with
larger cross-sections compared to
non-rotating stars (Romanova {\it et al.} 2001).
   Accelerated particles will propagate
most easily along the tail of the object and
may give the object an  elongated shape.
   An elongated shape is observed
around pulsar PSR 2224+65 in the
form of the Guitar
Nebulae (Cordes {\it et al.} 1993).
    Another
elongated pulsar trail was observed in
the X-ray band (Wang, Li \& Begelman 1993).
   This may be
connected with the stretching of magnetic field
lines by the ISM.

\section{Conclusions}

Axisymmetric MHD simulations of supersonic
motion of a star with an aligned
dipole magnetic field through the
ISM were performed for a wide range of
conditions.  We observed, that:

 1.~ The magnetized star acts an obstacle for the
flow of the ISM,
and a conical shock wave forms
as in the hydrodynamic case.

2.~ Long magnetotails form behind the star,
and reconnection is observed in the
tail.

3.~ In the $R_A\sim R_{acc}$ regime, some
matter accumulates around the star, but
most of the matter is deflected by
the magnetic field of the star
and flies away.
   The accretion rate to the star
is much smaller than that to a
non-magnetized star.

4.~ In the ``magnetic plow'' regime,  $R_A >>
R_{acc}$ and at high Mach numbers ${\cal
M}\sim 10 - 50$, no matter accumulation is
observed around the star.
   The density of matter in the
tail is very low.
Some matter accretes from the upwind pole.

5.~ When $R_A \gtrsim R_{acc}$, the
magnetic energy-density predominates in the
magnetotail.
   Part of this energy may radiate owing to
reconnection processes.
  The power is however small ($\sim 10^{21}
~{\rm erg/s}$ for typical parameters for
evolved pulsars and $\sim 10^{24} ~{\rm
erg/s}$ for magnetars), so that only the
closest magnetars may be possibly observed.
  For tail magnetic fields
$B \sim 10^{-4} - 10^{-6} ~{\rm G}$, the
tail flares or ``substorms'' may give
emission in the radio band.

6.~ Similar power to that
released by field reconnection is released in the
bow shock, which gives
radiation in the band from optical to X-ray.

7.~ Magnetic tails are expected to also form
in case of propagation of pulsars
through the ISM.
   In this case particles
accelerated by the pulsar will propagate
preferentially along the tail to give
an  elongated structure.

8.~ The presented simulations and
estimations can also be applied to
other magnetized stars propagating
through the ISM,
 such as magnetized white dwarfs, Ap stars,
and young stellar objects.

9.~ Propagation of magnetized
stars can lead to the
appearance of ordered magnetized structures
in the ISM.
   Also, these stars may give a contribution
to the magnetic flux of the Galaxy.

   The important influence of the rotation of the
star on the results described here
has been discussed briefly by
Romanova {\it et al.} (2001) and will
be treated thoroughly in a forthcoming
paper by our group.

\acknowledgments This work was supported in
part by NASA grant NAG5-9047, by NSF grant
AST-9986936, and by Russian program
``Astronomy.''
 RMM thanks NSF for a POWRE grant for partial
support. RVEL was partially supported by
grant NAG5-9735.
The authors thank Dr. V.V. Savelyev for providing with early
version of his numerical code, and thank Ira
Wasserman, Dave Chernoff, James Cordes and
Robert Duncan for valuable discussions.

\end{document}